\documentclass[showkeys,11pt]{revtex4} 

\usepackage[english]{babel} 
\usepackage{amssymb}
\usepackage{amsmath}
\usepackage{txfonts}
\usepackage{mathdots}
\usepackage[classicReIm]{kpfonts}
\usepackage[pdftex]{graphicx} 
\usepackage{multirow}
\usepackage{array}
\newcolumntype{P}[1]{>{\centering\arraybackslash}p{#1}}
\newcolumntype{M}[1]{>{\centering\arraybackslash}m{#1}}
\newcolumntype{N}{@{}m{0pt}@{}}

\begin{document}


\title{Multi-Scale Theory of Elasticity for Geomaterials}
\author{Christopher M. Szalwinski}
 \email{szalwin@yorku.ca}
 \affiliation{Lassonde School of Engineering, York University, Toronto ON M3J 1P3 Canada}

\begin{abstract}
The modern theory of elasticity and the first law of thermodynamics are cornerstones of engineering science that share the concept of reversibility. Engineering researchers have known for four decades that the modern theory violates the first law of thermodynamics when applied to the more commonly accepted empirical models of geomaterial stiffness. This paper develops a cross-scale theory of elasticity that is compatible with the empirical models and the first law of thermodynamics. This theory includes a material sample’s total-volume to solid-volume ratio as an independent internal variable, distinguishes deformation into uniform and contraction-swelling components, introduces a uniformity surface that partitions stress space into contraction and swelling sub-domains, couples the macroscopic properties to the volume ratio and extrapolates the accepted empirical models to states that include shear stress. This paper broadens the scope of the theory of elasticity to include soft condensed matter.
\end{abstract}

\keywords{constitutive relations; soft condensed matter; energy conservation; contraction-swelling; critical state soil mechanics}

\maketitle

\section{Introduction}

The states of repair of our countries' infrastructures reflect our theoretical understanding of earth materials. These materials include sands, gravel, gypsum, clay, shale, rock, and composites like glass, concrete, plaster, bricks, and asphalt [1]. They are economically essential to the global construction industry and react to external forces in complex ways. Geotechnical engineers, to meet design serviceability requirements for landfills, land surfaces, and land, sea and sub-surface structures, predict deformations well away from failure conditions. In their analyses, they rely on the modern theory of elasticity and expect closed loading cycles to conserve energy. However, for nearly half a century, researchers have claimed that implementing the more commonly accepted empirical models of soil deformation can violate the first law of thermodynamics [2-28]. This apparent violation of a well-established law based on experience is one example of a problem that involves crossing length scales.

Multi-scale investigations have been growing rapidly in materials science. The NSF Report on Simulation-Based Engineering Science [29] describes the transformation to multi-scale modeling and simulation as a powerful paradigm shift in engineering science, with disparities in cross-scale descriptions appearing in virtually all areas of science and engineering. The report refers to the ensemble of disparities as the tyranny of scales. These disparities focus attention on exploiting mesoscopic data to bridge the gaps between the top-down and bottom-up models of systems with neither strategy alone sufficing to yield the observable higher scale properties [30].

The history of materials science has shown us that a theory of elasticity that is based on a mesoscopic model alone is at best tentative. After a century-long contest, the multi-constancy tradition prevailed over the rari-constancy tradition [31]. The multi-constancy tradition is top-down, assumes that the superposition of pressures is due to a variety of displacements and defines pressures as linear functions of those displacements [32].  The rari-constancy tradition is bottom-up and models a body as composed of molecules with actions between them being in the line that joins the molecules [32]. The modern theory of elasticity is entirely within the former tradition.

The modern theory of elasticity requires two coefficients to describe a material that lacks directional preference (an isotropic material): the bulk modulus and the shear modulus. The bulk modulus specifies its stiffness in volumetric deformation; the shear modulus specifies its stiffness in distortion. The more commonly accepted empirical expressions for the bulk modulus of a soil sample are linear functions of effective pressure and specific volume [5,33-35] or linear functions of effective pressure alone [36,37]. Effective pressure is Cauchy pressure less interstitial fluid pressure. Specific volume is the ratio of a sample's volume to that of its solid constituents. The more commonly accepted empirical expressions for the shear modulus of a soil sample at small strains include a proper fractional power function of effective pressure and an improper fractional power function of specific volume [38-40].

Zytynski \textit{et al.}[2] demonstrated that these empirical models, which described bulk modulus as a linear function of effective pressure, are non-conservative. Although a classical conservative solution hosting bulk and shear moduli with \textit{identical} exponents for their power functions of pressure has been developed [41,42], no energetically conservative solution is available that supports the \textit{different} exponents for these two moduli evident in the empirical models.

Soils, powders, bulk solids, and other aggregates exhibit distinct material properties at macroscopic and mesoscopic scales. Their macroscopic stiffnesses and natural free states vary with packing. The modern theory of elasticity assumes uniform strain across all length scales and retains memory of a unique natural free state [43]. Each assumption is overly restrictive for these geomaterials.

This research paper develops a multi-scale theory of elasticity that includes specific volume as an independent internal state variable. The theory admits a continuum of natural free states, defines separate constitutive relations at macroscopic and mesoscopic scales, conserves energy across closed loading cycles and supports different exponents in the power functions of pressure for the bulk and shear moduli.

The body of this paper consists of 5 sections. Section 2 describes the mesoscopic model. It decomposes a representative element's deformation into uniform and differential parts. The differential part models deformation that involves a change in packing. Section 3 partitions stress space into contraction and swelling sub-domains and defines a contraction-swelling modulus that specifies the element's stiffness to a change in packing. Section 4 presents the internal energy potentials and the formal expressions for the macroscopic elasticity and compliance tensors and establishes their major symmetry. Section 5 derives two solutions for isotropic materials, one that highlights the theory's conceptual features and a more refined solution that is compatible with the empirical models accepted by geotechnical engineers. Section 6 reviews the published support in data for fine Ottawa sand, tire-derived aggregates, and select porous solids. This section concludes by comparing the theory to Critical State Soil Mechanics [5,33,35], proposing refinements to the latter and highlighting a fundamental difference at its limit.

\section{Mesoscopic Model}

Consider a geomaterial sample that consists of a large number of solid particles. The particles are in contact with one another and form the skeleton that defines the sample's boundary. The particles remain within the boundary, but and open to rearrangement; that is, the skeleton that defines the sample's boundary can change. The sample's pore content flows between the particles and can cross the sample's boundary.

The continuum element that represents this prototypical sample consists of a solid phase and an interstitial phase. The solid phase models the skeleton in its \textit{current} state; that is, the particles in their current arrangement. The interstitial phase models the pore content; that is, the fluid flowing through the solid phase and seeping across the element's boundary. The element's specific volume is the ratio of the sample's volume, $V$, to that of its solid particles, $V_s$:

\begin{tabular}{p{0.5in}p{3.4in}p{0.5in}}
 & $\nu \equiv V/V_s\ $  & (1) \\
\end{tabular}

\noindent
The element's porosity is the ratio of the sample's volume to its pore volume:

\begin{tabular}{p{0.5in}p{3.4in}p{0.5in}}
 & $\eta \equiv (V-V_s)/V=1-{\nu }^{-1}$ & (2) \\
\end{tabular}

\noindent
Specific volume and porosity are equivalent continuum measures of packing; both used in soil mechanics. Porosity is more common in the mechanics of porous solids. Specific volume and porosity are measurable but not directly controllable.

As the element's specific volume changes, so does each of its phases. The solid phase for one specific volume is distinct from the solid phase for any other specific volume. A change in specific volume involves a shift in the element's solid phase from that for the initial specific volume to that for the updated specific volume. These shifts represent rearrangements of the particles within the sample. The element model a sample with enough particles for all changes in its specific volume to appear to be continuous.

\subsection{Components of Volumetric Deformation}

\begin{figure}[ht!]
\centering
\includegraphics[scale=0.5]{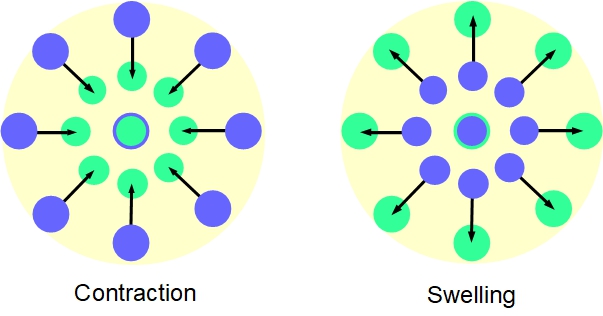}
\caption{Centric Deformation}
\label{fig:Fig1}
\end{figure}

The inclusion of specific volume, or porosity, as an independent variable enables a distinction between changes in average particle proximity and average particle radii; that is, an identification of two different aspects of sample deformation: intra-particle deformation and inter-particle deformation, with the latter measured relative to the former. Figure \ref{fig:Fig1} illustrates this degree of freedom. Note that the relative change in particle radii differs from the relative change in their proximity.

\begin{figure}[ht!]
\centering
\includegraphics[scale=0.5]{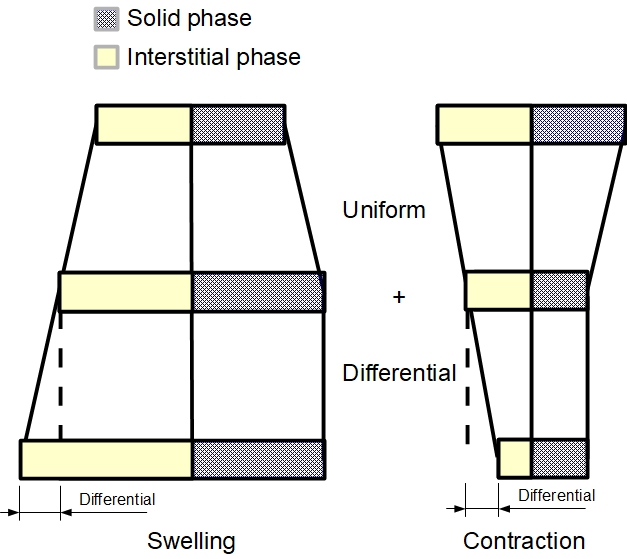}
\caption{Straining of Solid and Interstitial Phases}
\label{fig:Fig2}
\end{figure}

The representative element's volumetric deformation consists of a uniform component and a differential component. The uniform component models deformation at constant specific volume. Uniform deformation is identical for solid and interstitial phases. The differential component models the additional deformation of the interstitial phase. This part augments the uniform component of the interstitial phase and is directly related to the change in specific volume. Figure \ref{fig:Fig2} illustrates these two components.

Differential deformation is the centric part of the element's deformation (cf. the classical theory's rari-constancy model). It describes either contraction or swelling. \textit{Contraction} represents a decrease in the centroidal distances between adjacent particles. \textit{Swelling} represents an increase in the distances between adjacent particles. Contraction and swelling are each wholly distinct from uniform deformation.

\subsection{Packing Pressure}

To account for changes in energy associated with changes in packing, let us introduce a mesoscopic pressure within the element. This internal pressure is independent of the macroscopic pressure applied to the element's boundary.

Consider a sphere centered at the element's mass center as shown in Figure \ref{fig::Fig3}. Its surface represents the average mass centers of particles equidistant from the sample's mass center. As the element's specific volume changes the radius of the sphere changes; that is, the distances of the particles from the mass center change. Let us define the \textit{packing pressure} within the element as the mesoscopic pressure that maintains the sphere at its current radius and denote this pressure by $\phi $.

\begin{figure}[ht!]
\centering
\includegraphics[scale=0.5]{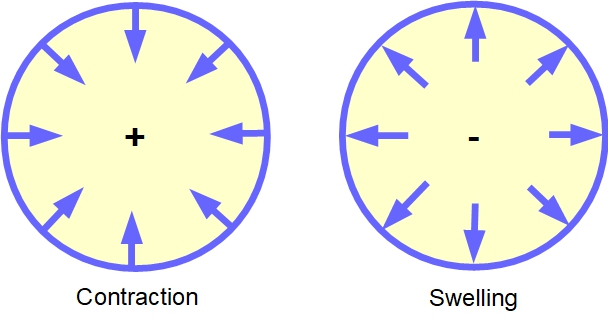}
\caption{Changes in Packing Pressure}
\label{fig:Fig3}
\end{figure}

Packing pressure can change even if the pressure applied at the boundary remains constant. Conversely, the pressure applied at the boundary can change even if the packing pressure remains constant. Consolidation is an example of a process that involves changes in internal pressure but does not necessarily involve any change in externally applied pressure.

A change in packing pressure either reduces or increases the sphere's radius. Contractive changes reduce its radius. Swelling changes increase its radius.

The extent of the change in the sphere's radius due to a change in packing pressure depends on the material's properties. Under the principle of local state [44], a change in packing pressure is related to the local change in specific volume but not to its gradient. The packing pressure at which specific volume remains unchanged is the element's current \textit{equilibrium packing pressure}. Assuming that an equilibrium packing pressure exists for each specific volume, let us define a \textit{contraction-swelling} curve in $\phi -\nu $ space that relates equilibrium packing pressures to specific volumes throughout the practical range of specific volumes:

\begin{tabular}{p{0.5in}p{3.4in}p{0.5in}}
 & $\beta \equiv \beta \left(\phi ,\nu \right)=0$ & (3) \\
\end{tabular}

\begin{figure}[ht!]
\centering
\includegraphics[scale=0.5]{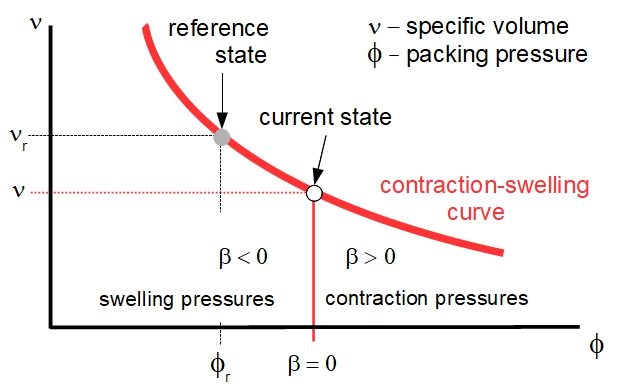}
\caption{Contraction-Swelling Constitutive Relation}
\label{fig:Fig4}
\end{figure}

\noindent
At lower pressures, the element's specific volume is highly sensitive to small changes in packing pressure; that is, the packing of the sample's particles can change significantly. On the other hand, at higher pressures, the element's specific volume is relatively insensitive to large changes in packing pressure. These two limiting conditions determine the general form of the contraction-swelling curve. This curve is illustrated in Figure \ref{fig:Fig4}. ${\phi }_r$ is an arbitrarily selected reference packing pressure for the element and ${\nu }_r$ is the specific volume corresponding to that pressure.

The subsets of contraction and swelling pressures for the element depend on its current specific volume. The current specific volume or its equilibrium packing pressure partitions the contraction-swelling curve into contraction and swelling segments. The current contraction pressures are the packing pressures that satisfy $\betaup$$\mathrm{>}$0. The current swelling pressures are those that satisfy $\betaup$$\mathrm{<}$0.

\subsection{Packing Energy}

Any work done by the packing pressure during a change in specific volume changes the element's packing energy. This work excludes all work involving uniform deformation; that is, all work done by externally applied stress in uniform deformation.

Considering the interstitial phase of the element as a sphere of radius $r$, the work done during a change in its radius is the work done by the packing pressure at the sphere's surface:

\begin{tabular}{p{0.5in}p{3.4in}p{0.5in}}
 & $\delta W=\ -\ \phi \ 4\pi r^2\delta r$ & (4) \\
\end{tabular}

\noindent
where $\delta $ denotes \textit{increment of}. The minus sign associates positive work with contraction. The change in the sphere's radius is directly related to the change in the element's specific volume:

\begin{tabular}{p{0.5in}p{2.65in}p{1.25in}}
 & $\delta \nu =\delta V/V =\ 3 \delta r/r$ & $for\ \delta V_s = 0$ (5) \\
\end{tabular}

\noindent
The change in the element's packing energy is the work done per unit volume:

\begin{tabular}{p{0.5in}p{3.4in}p{0.5in}}
 & $\delta P=\delta W\ /\ V$   & (6) \\
\end{tabular}

\noindent
where $P$ denotes \textit{packing energy}. From Eqs. (4), (5) and (6):

\begin{tabular}{p{0.5in}p{3.4in}p{0.5in}}
 & $\delta P=\ -\ \phi \ \delta \nu $ & (7) \\
\end{tabular}

\noindent
The element's packing energy follows from integration:

\begin{tabular}{p{0.5in}p{3.4in}p{0.5in}}
 & $P=\int^{\nu }_{{\nu }_r}{\delta P}$ & (8) \\
\end{tabular}

\noindent
Packing energy vanishes at the selected reference state (${\phi }_r,{\nu }_r$).

The appendix contains derivations of expressions for two packing energy potentials based on separate contraction-swelling constitutive relations.

\section{Macroscopic Model}

The element's specific volume and the externally applied stress define its state completely. A change in the applied stress may cause a change in specific volume and that change depends on the element's properties.

To identify the form of the relation between a change in the element's strain and any change in its specific volume consider a linearly elastic material with a compliance that varies with specific volume alone. The strain tensor, $\boldsymbol{\epsilon }$, for such a material, is the inner product of its compliance tensor, $\boldsymbol{C}(\nu )$, and the applied stress, $\boldsymbol{\sigma }$:

\begin{tabular}{p{0.5in}p{3.4in}p{0.5in}}
 & $\boldsymbol{\epsilon }=\boldsymbol{C}\left(\nu \right):\boldsymbol{\sigma }$ & (9) \\
\end{tabular}

\noindent
where \textbf{:} denotes \textit{inner tensor product} on two subscripts. The change in this strain tensor depends on both the change in the stress tensor and the change in specific volume. Differentiating Eq. (9) (cf. [45,46]) yields

\begin{tabular}{p{0.5in}p{3.4in}p{0.5in}}
 & $\delta \boldsymbol{\epsilon }=\boldsymbol{C}(\nu ):\delta \boldsymbol{\sigma }+\ \boldsymbol{\sigma }:(\partial \boldsymbol{C}(\nu )/\partial \nu ) \delta \nu $ & (10) \\
\end{tabular}

\noindent
The first term on the right-hand side is the \textit{uniform} contribution to the strain increment. This contribution models identical straining of the solid and interstitial phases; that is, straining at constant specific volume. The second term is the \textit{differential} contribution. It models the change in specific volume at constant stress.

Given this decomposition (Eq. (10)), consider a more complex material with a compliance that also varies with externally applied stress. The strain increment tensor consists of uniform and differential components:

\begin{tabular}{p{0.5in}p{3.4in}p{0.5in}}
 & $\delta \boldsymbol{\epsilon }=\ \delta {\boldsymbol{\epsilon }}_u+\ \delta {\boldsymbol{\epsilon }}_d$ & (11) \\
\end{tabular}

\noindent
where subscripts $u$ and $d$ denote \textit{uniform} and \textit{differential} respectively. The uniform component is linearly related to the stress increment tensor:

\begin{tabular}{p{0.5in}p{3.4in}p{0.5in}}
 & $\delta {\boldsymbol{\epsilon }}_u={\boldsymbol{C}}_u\left(\boldsymbol{\sigma },\nu \right):\delta \boldsymbol{\sigma }$ & (12) \\
\end{tabular}

\noindent
where ${\boldsymbol{C}}_u$ denotes the \textit{uniform compliance tensor}. This component pervades all length scales. The differential component is given by

\begin{tabular}{p{0.5in}p{3.4in}p{0.5in}}
 & $\delta {\boldsymbol{\epsilon }}_d=\boldsymbol{\omega }\left(\boldsymbol{\sigma },\nu \right)\ \delta \mu $ & (13) \\
\end{tabular}

\noindent
where $\boldsymbol{\omega }$ denotes the \textit{normalized coupling tensor} [45]. The directions of this component depend on the current state. The scalar multiplier, $\delta \mu $, is its magnitude. It is positive-valued in contraction and negative-valued in swelling. Its relation to the change in specific volume is established in sub-section \ref{sec::comela} below (Eq. (53)).

\begin{figure}[ht!]
\centering
\includegraphics[scale=0.5]{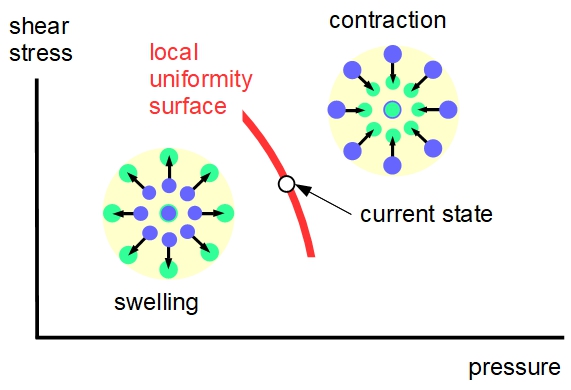}
\caption{Uniformity Surface through the Current State}
\label{fig:Fig5}
\end{figure}

The magnitude of the differential component depends not only on the element's state but also on the applied stress increment. To distinguish between contraction and swelling processes, consider a local surface in stress sub-space that passes through the current stress state and partitions the neighboring sub-space into contraction and swelling sub-domains as illustrated in Figure \ref{fig:Fig5}. Let us represent this surface by

\begin{tabular}{p{0.5in}p{3.4in}p{0.5in}}
 & $b \equiv b\left(\boldsymbol{\sigma },\nu \right)=0$ & (14) \\
\end{tabular}

\noindent
Contraction states satisfy $b$$\mathrm{>}$0, while swelling states satisfy $b$$\mathrm{<}$0. Only stress increments along the surface produce purely uniform straining. Let us call this surface the element's \textit{uniformity surface}. If ${\boldsymbol{n}}$ denotes its normalized gradient:

\begin{tabular}{p{0.5in}p{3.4in}p{0.5in}}
 & $\boldsymbol{n} \equiv \frac{\partial b}{\partial \boldsymbol{\sigma }}\ /\parallel \frac{\partial b}{\partial \boldsymbol{\sigma }}\parallel \boldsymbol{\mathrm{\ }}\boldsymbol{\mathrm{\ }}$ & (15) \\
\end{tabular}

\noindent
then the stress increments that preserve the element's specific volume are normal to this gradient:

\begin{tabular}{p{0.5in}p{2.65in}p{1.25in}}
 & ${\boldsymbol{n}}:\delta \boldsymbol{\sigma }=0$ & $for\ \delta \mu = 0$ (16) \\
\end{tabular}

\noindent
$\parallel \boldsymbol{x}\parallel $ denotes the positive-valued magnitude of $\boldsymbol{x}$.

Since stress increments tangential to the surface do not cause any differential straining, only that component of the stress increment that is normal to the surface change specific volume. That is, the signed magnitude of the differential component of the strain increment tensor is linearly related to the component of the stress increment tensor that is normal to the local uniformity surface. This consistency relation may be written as

\begin{tabular}{p{0.5in}p{3.4in}p{0.5in}}
 & $\boldsymbol{n}:\delta \boldsymbol{\sigma }-S\delta \mu =0$ & (17) \\
\end{tabular}

\noindent
where $S$ denotes the \textit{contraction-swelling modulus}. The expression for the signed magnitude follows directly from this relation:

\begin{tabular}{p{0.5in}p{3.4in}p{0.5in}}
 & $\delta \mu \ \mathrm{=} \ \boldsymbol{n}:\delta \boldsymbol{\sigma }/S$ & (18) \\
\end{tabular}

\noindent
The contraction-swelling modulus specifies the element's stiffness to differential deformation independently of its stiffness to uniform deformation. Aggregates, which exhibit noticeable changes in packing, have low-valued moduli. Porous solids, which exhibit minor changes in packing, have relatively high-valued moduli.

Not all processes in this model are equilibrium processes. Stress increments directed along the local uniformity surface preserve specific volume and maintain multi-scale equilibrium. Since the specific volume does not change, the equilibrium packing pressure remains constant. Stress increments directed off the surface initiate non-equilibrium processes. Internal adjustments in specific volume may occur at different rates than the rates of change of external macroscopic stresses. While processes to states along the surface are unconstrained, processes to states off the surface progress from initially constrained states to ultimately unconstrained states. At multi-scale equilibrium, the end state satisfies macroscopic equilibrium and the packing pressure is the equilibrium packing pressure for the end state's specific volume. That is, the end stress state lies on the local uniformity surface for the end stress state and the end specific volume.

This approach has some precedents. Augmenting thermodynamic solutions with internal state variables is well-established practice [47]. Internal state variables support constraint modeling of viscoelasticity and relaxation near equilibrium [48]. Constrained equilibrium modeling extends to the averaging of an internal variable [49]. Partitioning stress sub-space into sub-domains of differing responses is a feature of the mathematical theory of elasto-plasticity, as is a consistency relation.

\subsection{Compliance and Elasticity}
\label{sec::comela}

The macroscopic compliance tensor for the element models all processes, regardless of the stress increment tensor directions and predicts the strain increment based on the applied stress increment. Substituting Eqs. (12), (13) and (18) into Eq. (11) yields

\begin{tabular}{p{0.5in}p{3.4in}p{0.5in}}
 & $\delta \boldsymbol{\epsilon }\  \mathrm{=}\  \boldsymbol{C}\boldsymbol{:}\delta \boldsymbol{\sigma }$ & (19) \\
\end{tabular}

\noindent
where $\boldsymbol{C}$ denotes the element's \textit{macroscopic compliance tensor}:

\begin{tabular}{p{0.5in}p{3.4in}p{0.5in}}
 & $\boldsymbol{C} \equiv {\boldsymbol{\ }\boldsymbol{C}}_{\boldsymbol{u}}\boldsymbol{+}\boldsymbol{\omega }\boldsymbol{\otimes }\boldsymbol{n}/S$ & (20) \\
\end{tabular}

\noindent
where $\otimes$ denotes \textit{outer tensor product}. The uniform compliance tensor predicts the element's compliance to unconstrained change; that is, for stress increments along the local uniformity surface. The rightmost term describes the added strain increment as specific volume progresses from its initially constrained value to its multi-scale equilibrium value.

The macroscopic elasticity tensor predicts the stress increment tensor corresponding to an applied strain increment tensor. Substituting Eq. (12) into Eq. (11), inverting the result and substituting Eq. (13) yields

\begin{tabular}{p{0.5in}p{3.4in}p{0.5in}}
 & $\delta \boldsymbol{\sigma }\ \mathrm{=}\ {\boldsymbol{E}}_u\boldsymbol{:}\boldsymbol{\delta }\boldsymbol{\epsilon }-{\boldsymbol{E}}_u\boldsymbol{:}\boldsymbol{\omega }\ \delta \mu $ & (21) \\
\end{tabular}

\noindent
where ${\boldsymbol{E}}_u$ denotes the \textit{uniform elasticity tensor}:

\begin{tabular}{p{0.5in}p{3.4in}p{0.5in}}
 & ${\boldsymbol{E}}_u\ = {\boldsymbol{C}}^{-1}_u$ & (22) \\
\end{tabular}

\noindent
The expression for the signed magnitude of the strain increment's differential component follows from the consistency relation. Substituting Eq. (21) into Eq. (17) yields

\begin{tabular}{p{0.5in}p{3.4in}p{0.5in}}
 & $\delta \mu = (\boldsymbol{n}\ {\boldsymbol{:}\boldsymbol{E}}_u\boldsymbol{:}\boldsymbol{\delta }\epsilon  )\ /\ (S+\boldsymbol{n}:{\boldsymbol{E}}_u\boldsymbol{:}\boldsymbol{\omega })$ & (23) \\
\end{tabular}

\noindent
Substituting Eq. (23) into Eq. (21) yields

\begin{tabular}{p{0.5in}p{3.4in}p{0.5in}}
 & $\delta \boldsymbol{\sigma } = \boldsymbol{E}\boldsymbol{:}\boldsymbol{\delta }\boldsymbol{\epsilon }$ & (24) \\
\end{tabular}

\noindent
where $\boldsymbol{E}$ denotes the element's \textit{macroscopic elasticity tensor}:

\begin{tabular}{p{0.5in}p{3.4in}p{0.5in}}
 & $\boldsymbol{E} \equiv \boldsymbol{\ }\boldsymbol{E}_u-({\boldsymbol{E}}_u \boldsymbol{:}\boldsymbol{\omega }\boldsymbol{\otimes }{\boldsymbol{n}\boldsymbol{:}\boldsymbol{E}}_u)\ /\ (S+\boldsymbol{n}\boldsymbol{:}{\boldsymbol{E}}_u:\boldsymbol{\omega })$ & (25) \\
\end{tabular}

\noindent
The rightmost term in Eq. (25) relaxes the uniform stiffness accounting for the added freedom as specific volume changes from its initially constrained value to its unconstrained end value, at multi-scale equilibrium.

The relation between the normalized coupling tensor ($\boldsymbol{\omega }$) and the normalized gradient to the uniformity surface ($\boldsymbol{n}$) in the expressions for both macroscopic tensors is established in sub-section \ref{sec::majsym} below.

\subsection{Contraction-Swelling Modulus}

The contraction-swelling modulus specifies the element's stiffness to changes in differential deformation regardless of uniform deformation. Its value can be estimated from measurements of volumetric and distortional stiffness at isotropic loading states.

The pressure or mean-normal stress invariant is defined as

\begin{tabular}{p{0.5in}p{3.4in}p{0.5in}}
 & $p\ \equiv \boldsymbol{\sigma }:\boldsymbol{I}/3$ & (26) \\
\end{tabular}

\noindent
where $\boldsymbol{I}$ denotes the identity tensor. Let us assume that the normalized coupling tensor and the normalized gradient to the uniformity surface are identity transformations at all isotropic loading states:

\begin{tabular}{p{0.5in}p{2.65in}p{1.25in}}
 & \textbf{                       }$\boldsymbol{\omega } = \boldsymbol{n} = \boldsymbol{I}$ &      $for\ \boldsymbol{\sigma }=p_o\boldsymbol{I}$ (27) \\
\end{tabular}

\noindent
where $p_o\ $denotes the applied pressure at any isotropic state. The volumetric strain increment is defined as

\begin{tabular}{p{0.5in}p{3.4in}p{0.5in}}
 & $\delta \epsilon \equiv \boldsymbol{I}:\delta \boldsymbol{\epsilon }$ & (28) \\
\end{tabular}

\noindent
The macroscopic bulk modulus, $K$, relates this invariant linearly to the pressure increment:

\begin{tabular}{p{0.5in}p{2.65in}p{1.25in}}
 &                  $\delta \epsilon =K^{-1}\ \delta p_o$ & $for\ \boldsymbol{\sigma }=p_o\boldsymbol{I}$ (29) \\
\end{tabular}

\noindent
$K$ is the tangential slope of the unloading-reloading line in $\epsilon -p_o\ $space.

Substituting Eq. (27) into Eq. (25) and contracting twice yields:

\begin{tabular}{p{0.5in}p{2.65in}p{1.25in}}
 &                    $K^{-1} =K^{-1}_u+S^{-1}$ & $for\ \boldsymbol{\sigma }=p_o\boldsymbol{I}$ (30) \\
\end{tabular}

\noindent
where $K_u$ denotes the \textit{uniform bulk modulus} of the element:

\begin{tabular}{p{0.5in}p{3.4in}p{0.5in}}
 & $K_u\ \equiv  \boldsymbol{I}:{\boldsymbol{E}}_u:\boldsymbol{I}\ /\ 9$ & (31) \\
\end{tabular}

\noindent
The expression for the contraction-swelling modulus is:

\begin{tabular}{p{0.5in}p{3.4in}p{0.5in}}
 & $S={\left[K^{-1}-K^{-1}_u\ \right]}^{-1}={\left[\left(\frac{\delta \epsilon }{\delta p_o}\right)-K^{-1}_u\ \right]}^{-1}$ & (32) \\
\end{tabular}

\noindent
The value of the uniform bulk modulus can be estimated from an empirically determined uniform shear compliance by assuming a constant Poisson's ratio across the family of solid phases of all specific volumes.

If the element is significantly more compliant than its solid phase, a first approximation for the volumetric strain increment is

\begin{tabular}{p{0.5in}p{2.65in}p{1.25in}}
 & $\delta \epsilon \ \approx \ -\ \delta \nu /\nu $ & $for\ K_u\gg K$ (33) \\
\end{tabular}

\noindent
In this case, the expression for the contraction-swelling modulus reduces to

\begin{tabular}{p{0.5in}p{2.65in}p{1.25in}}
 & $S\approx \delta p_o/\delta \epsilon $ & $for\ K_u\gg K$ (34) \\
\end{tabular}

\section{Internal Energy}

The internal energy of the element integrates the macroscopic and mesoscopic models. A sufficient condition for conservation of internal energy in a closed process is the existence of a potential in the independent state variables.

In a kinematic description, strain and specific volume are the independent state variables, the internal energy potential is the scalar measure of the element's state and stress is the dependent state variable. In a kinetic description, stress and specific volume are the independent state variables, the complementary internal energy potential is the scalar measure of state and strain is the dependent variable. The expressions for the stress, strain, packing pressure, elasticity, compliance and coupling tensors follow from these two potentials.

\subsection{Potential Functions}

The internal energy potential,$\ E\left(\boldsymbol{\epsilon },\nu \right)$, describes the element's physical state in terms of its strain and specific volume. The potential's uniform version, $U\left(\boldsymbol{\epsilon },\nu \right)$, which describes its physical state for a prescribed specific volume, is the difference between the internal energy potential, $E\boldsymbol{(}\boldsymbol{\epsilon },\nu )$\textbf{,} and the element's packing energy, $P(\nu )$:

\begin{tabular}{p{0.5in}p{3.4in}p{0.5in}}
 & $U\left(\boldsymbol{\epsilon },\bar{\nu } \right)=E\left(\boldsymbol{\epsilon },\nu \right)\ -\ P\boldsymbol{(}\nu )$ & (35) \\
\end{tabular}

\begin{figure}[ht!]
\centering
\includegraphics[scale=0.5]{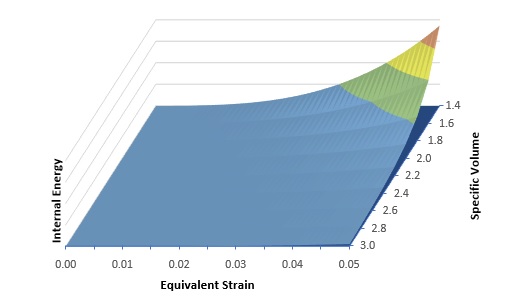}
\caption{Internal Energy}
\label{fig:Fig6}
\end{figure}

\noindent
$U\left(\boldsymbol{0 },\bar{\nu } \right)$ represents the natural free state for the prescribed specific volume ($\nu = \bar{\nu }$). Figure \ref{fig:Fig6} illustrates the internal energy potential surface as a function of equivalent strain and specific-volume. The reference state is the state at which the specific volume is the reference specific volume ($\nu ={\nu }_r$) and all strain vanishes ($E\left(\boldsymbol{0},{\nu }_r\right)$).

The complementary internal energy potential, $C\left(\boldsymbol{\sigma },\nu \right)$, describes the energy that the element can transfer to its environment. This potential's uniform version, $Q\left(\boldsymbol{\epsilon },\bar{\nu } \right)$, which describes the transferable energy for a prescribed specific volume, is the sum of the complementary internal energy, $C\left(\boldsymbol{\sigma },\nu \right)$, and the element's packing energy, $P\left(\nu \right)$:

\begin{tabular}{p{0.5in}p{3.4in}p{0.5in}}
 & $Q\left(\boldsymbol{\sigma },\bar{\nu } \right)=C\left(\boldsymbol{\sigma },\nu \right)+P\left(\nu \right)$ & (36) \\
\end{tabular}

\begin{figure}[ht!]
\centering
\includegraphics[scale=0.5]{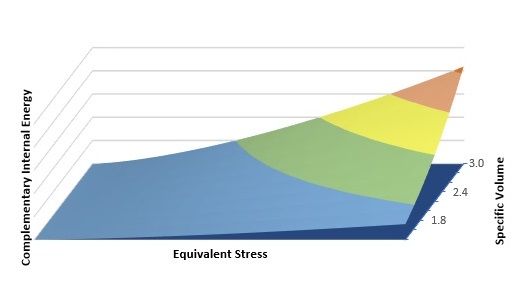}
\caption{Complementary Internal Energy}
\label{fig:Fig7}
\end{figure}

\noindent
Figure \ref{fig:Fig7} illustrates the complementary internal energy potential surface as a function of stress and specific volume. The reference state for this surface is the state at which the specific volume is the reference specific volume ($\nu ={\nu }_r$) and stress vanishes ($C\left(\boldsymbol{0},{\nu }_r\right)$).

The uniform complementary internal energy is the partial Legendre transform of element's uniform internal energy with respect to strain [50]:

\begin{tabular}{p{0.5in}p{3.4in}p{0.5in}}
 & $U\left(\boldsymbol{\epsilon },\nu \right)+\ Q\left(\boldsymbol{\sigma },\nu \right)=\boldsymbol{\sigma }:\boldsymbol{\epsilon }$ & (37) \\
\end{tabular}

\noindent
Changes in packing energy are passive and its terms in Eqs. (35) and (36) cancel one another in Eq. (37). The uniform potentials are the energy potentials of the modern theory.

\subsection{Stress, Strain, Elasticity, Compliance and Coupling Tensors}
\label{sec:strstr}

The stress, strain and packing pressure ($\boldsymbol{\sigma },\ \boldsymbol{\epsilon },\ \phi $) are partial derivatives of the internal energy and complementary internal energy potential functions.

Differentiating Eq. (35) and substituting Eq. (7) into the result distinguishes an internal energy increment into macroscopic and mesoscopic components:

\begin{tabular}{p{0.5in}p{3.4in}p{0.5in}}
 & $\delta U={\left(\frac{\partial E}{\partial \boldsymbol{\epsilon }}\right)}_{\nu }:\delta \boldsymbol{\epsilon }+{\left(\frac{\partial E}{\partial \nu }\right)}_{\boldsymbol{\epsilon }}\ \delta \nu +\phi \ \delta \nu $ & (38) \\
\end{tabular}

\noindent
At multi-scale equilibrium, specific volume is constant, and the work done by the stress on the element is

\begin{tabular}{p{0.5in}p{3.4in}p{0.5in}}
 & $\boldsymbol{\sigma }:\delta \boldsymbol{\epsilon } = {\left(\frac{\boldsymbol{\partial }E}{\boldsymbol{\partial }\boldsymbol{\epsilon }}\right)}_{\nu }:\boldsymbol{\delta }\boldsymbol{\epsilon }$ & (39) \\
\end{tabular}

\noindent
Equating Eqs. (38) and (39) gives the expressions for the stress and the packing pressure:

\begin{tabular}{p{0.5in}p{3.4in}p{0.5in}}
 & $\boldsymbol{\sigma }={\left(\frac{\partial E}{\partial \boldsymbol{\epsilon }}\right)}_{\nu }$ & (40) \\
 & $\phi = -\ {\left(\frac{\partial E}{\partial \nu }\right)}_{\boldsymbol{\epsilon }}$ & (41) \\
\end{tabular}

\noindent
Differentiating Eq. (36) and substituting Eq. (7) into the result distinguishes the complementary internal energy increment into macroscopic and mesoscopic components:

\begin{tabular}{p{0.5in}p{3.4in}p{0.5in}}
 & $\delta Q={\left(\frac{\partial C}{\partial \boldsymbol{\sigma }}\right)}_{\nu }:\delta \boldsymbol{\sigma }+{\left(\frac{\partial C}{\partial \nu }\right)}_{\boldsymbol{\sigma }}\delta \nu -\ \phi \ \delta \nu $ & (42) \\
\end{tabular}

\noindent
At multi-scale equilibrium, specific volume is constant, and the complementary work done is given by

\begin{tabular}{p{0.5in}p{3.4in}p{0.5in}}
 & $\boldsymbol{\epsilon }:\delta \boldsymbol{\sigma }={\left(\frac{\partial C}{\partial \boldsymbol{\sigma }}\right)}_{\nu }:\boldsymbol{\delta }\boldsymbol{\sigma }$ & (43) \\
\end{tabular}

\noindent
Equating Eqs. (42) and (43) gives the expressions for the total strain and the packing pressure:

\begin{tabular}{p{0.5in}p{3.4in}p{0.5in}}
 & $\boldsymbol{\epsilon }={\left(\frac{\partial C}{\partial \boldsymbol{\sigma }}\right)}_{\nu }$ & (44) \\
 & $\phi =\ {\left(\frac{\partial C}{\partial \nu }\right)}_{\boldsymbol{\sigma }}$ & (45) \\
\end{tabular}

\noindent
Eqs. (41) and (45) are equivalent expressions for packing pressure.

The uniform elasticity, uniform compliance and coupling tensors are second derivatives of the energy potentials. Differentiating Eq. (40) distinguishes the stress increment into uniform and differential components:

\begin{tabular}{p{0.5in}p{3.4in}p{0.5in}}
 & $\delta \boldsymbol{\sigma }={\boldsymbol{E}}_u:\delta \boldsymbol{\epsilon }+\frac{{\partial }^2E}{\partial \nu \partial \boldsymbol{\epsilon }}\ \delta \nu $ & (46) \\
\end{tabular}

\noindent
where ${\boldsymbol{E}}_u$ denotes the uniform elasticity tensor:

\begin{tabular}{p{0.5in}p{3.4in}p{0.5in}}
 & ${\boldsymbol{E}}_u \equiv \frac{{\boldsymbol{\partial }}^{\boldsymbol{2}}E}{\partial \boldsymbol{\epsilon }\partial \boldsymbol{\epsilon }}$ & (47) \\
\end{tabular}

\noindent
Differentiating Eq. (44) distinguishes the strain increment into uniform and differential components:

\begin{tabular}{p{0.5in}p{3.4in}p{0.5in}}
 & $\delta \boldsymbol{\epsilon }={\boldsymbol{C}}_u:\delta \boldsymbol{\sigma }+\boldsymbol{\mathit{\Omega}}\ \delta \nu $ & (48) \\
\end{tabular}

\noindent
where $\boldsymbol{\mathit{\Omega}}$ denotes the coupling tensor. The uniform compliance and coupling tensors are second partial derivatives of the complementary internal energy potential:

\begin{tabular}{p{0.5in}p{3.4in}p{0.5in}}
 & ${\boldsymbol{C}}_u = \frac{{\boldsymbol{\partial }}^{\boldsymbol{2}}C}{\partial \boldsymbol{\sigma }\partial \boldsymbol{\sigma }}$ & (49) \\
 & $\boldsymbol{\mathit{\Omega}}=\frac{{\partial }^2C}{\partial \nu \partial \boldsymbol{\sigma }}$ & (50) \\
\end{tabular}

\noindent
${\boldsymbol{E}}_u$ and ${\boldsymbol{C}}_u$ are inverses of one another.

Normalizing the coupling tensor and comparing Eq. (48) to Eq. (11) with Eq. (12) identifies the differential component of strain increment tensor as the product of the coupling tensor and the specific volume increment:

\begin{tabular}{p{0.5in}p{3.4in}p{0.5in}}
 & $\boldsymbol{\delta }{\boldsymbol{\epsilon }}_d = -\ \boldsymbol{\mathit{\Omega}}\ \delta \nu $ & (51) \\
\end{tabular}

\noindent
Substituting Eq. (51) into Eq. (13) relates the coupling tensor to the normalized coupling tensor ($\boldsymbol{\mathit{\omega }} $) and the signed magnitude of the strain increment's differential component ($\delta \mu $) to the specific volume increment:

\begin{tabular}{p{0.5in}p{3.4in}p{0.5in}}
 & $\boldsymbol{\omega } = \boldsymbol{\mathit{\Omega}}\ /\parallel \boldsymbol{\mathit{\Omega}}\parallel $ & (52) \\
 & $\delta \mu \ \mathrm{=}\ - \parallel \boldsymbol{\mathit{\Omega}}\parallel \ \delta \nu $ & (53) \\
\end{tabular}

\noindent
The minus sign indicates that a specific volume decrement ($\delta \nu <0$) is contractive ($\delta \mu >0$). The material properties that determine differential deformation enter entirely through the coupling tensor.

\subsection{Major Symmetry}
\label{sec::majsym}

The macroscopic analysis of sub-section 3.1 expresses the compliance and elasticity tensors in terms of the normalized gradient to the uniformity surface and the normalized coupling tensor. Any stress increment directed off the uniformity surface for the current state changes the element's specific volume. During the process, the element's stress state and specific volume may change asynchronously. The major symmetry of the compliance and elasticity tensors follows from multi-scale equilibrium.

For the element to be in a state of multi-scale equilibrium, its stress state must lie on the local uniformity surface for the current specific volume and its packing pressure must be the equilibrium packing pressure for that same specific volume:

\begin{tabular}{p{0.5in}p{3.4in}p{0.5in}}
 & $b\left(\boldsymbol{\sigma },\nu \right)=\ \beta \left(\phi ,\nu \right)=0$ & (54) \\
\end{tabular}

\noindent
Stress increments along the uniformity surface maintain multi-scale equilibrium. Differentiating Eq. (54) holding specific volume constant relates the surface gradient to the slope of the contraction-swelling curve through the packing pressure gradient:

\begin{tabular}{p{0.5in}p{3.4in}p{0.5in}}
 & $\frac{\partial b}{\partial \boldsymbol{\sigma }}=\frac{\partial \beta }{\partial \phi \ }\cdot \frac{\partial \phi }{\partial \boldsymbol{\sigma }}$ & (55) \\
\end{tabular}

\noindent
Differentiating Eq. (45) holding specific volume constant yields the expression for the packing pressure gradient:

\begin{tabular}{p{0.5in}p{3.4in}p{0.5in}}
 & $\frac{\partial \phi }{\partial \boldsymbol{\sigma }}=\frac{{\partial }^2C}{\partial \boldsymbol{\sigma }\partial \nu }$ & (56) \\
\end{tabular}

\noindent
Identity of cross derivatives of the complementary strain energy function (a Maxwell relation) relates this gradient to the coupling tensor:

\begin{tabular}{p{0.5in}p{3.4in}p{0.5in}}
 & $\frac{\partial \phi }{\partial \boldsymbol{\sigma }}=\ \frac{{\partial }^2C}{\partial \nu \partial \boldsymbol{\sigma }}=\ \boldsymbol{\mathit{\Omega}}$ & (57) \\
\end{tabular}

\noindent
Substituting Eq. (57) into Eq. (55) relates the uniformity surface gradient to the slope of the contraction-swelling curve through the coupling tensor:

\begin{tabular}{p{0.5in}p{3.4in}p{0.5in}}
 & $\frac{\partial b}{\partial \boldsymbol{\sigma }}=\ \boldsymbol{\mathit{\Omega}}\ \frac{\partial \beta }{\partial \phi \ }$ & (58) \\
\end{tabular}

\noindent
Normalizing the surface gradient and substituting Eq. (52) yields

\begin{tabular}{p{0.5in}p{3.4in}p{0.5in}}
 & $\boldsymbol{n} = \boldsymbol{\omega }\parallel \boldsymbol{\mathit{\Omega}}\parallel \left(\frac{\partial \beta }{\partial \phi }\right)/\parallel \frac{\partial b}{\partial \boldsymbol{\sigma }}\parallel $ & (59) \\
\end{tabular}

\noindent
Since both $\boldsymbol{n}$ and $\boldsymbol{\omega }$ are normalized,

\begin{tabular}{p{0.5in}p{3.4in}p{0.5in}}
 & $\boldsymbol{n}\ =\ \boldsymbol{\omega }$ & (60) \\
 & $\parallel \frac{\partial b}{\partial \boldsymbol{\sigma }}\parallel \ =\ \parallel \boldsymbol{\mathit{\Omega}}\parallel \left(\frac{\partial \beta }{\partial \phi \ }\right)$ & (61) \\
\end{tabular}

\noindent
The directions of the coupling tensors are independent of the contraction-swelling relation.

Substituting Eq. (60) into Eqs. (20) and (25) simplifies the expressions for the macroscopic compliance and elasticity tensors:

\begin{tabular}{p{0.5in}p{3.4in}p{0.5in}}
 & $\boldsymbol{C} = {\boldsymbol{C}}_u+{\boldsymbol{\omega }\boldsymbol{\otimes }\boldsymbol{\omega }}\ /\ {S}$ & (62) \\
 & $\boldsymbol{E} = \boldsymbol{E}_u-({\boldsymbol{E}}_u\boldsymbol{:}\boldsymbol{\omega }\boldsymbol{\otimes }\boldsymbol{\omega }\boldsymbol{:}{\boldsymbol{E}}_u)\ /\ (S+\ \boldsymbol{\omega }\boldsymbol{\ }\boldsymbol{:}{\boldsymbol{E}}_u\boldsymbol{:}\boldsymbol{\omega })$ & (63) \\
\end{tabular}

\noindent
That is, a sufficient condition for major symmetry of each tensor is multi-scale equilibrium (Eq. 55).

\section{Isotropic Materials}

The generally accepted empirical expressions for bulk and shear moduli of soil samples have been established for isotropic materials at isotropic states. The theory's linear specialization for these materials demonstrates the simplest resolution of the energy conservation issue for these materials. Its non-linear specialization demonstrates resolution of the differing exponents issue in the power functions of the bulk and shear moduli.

\subsection{General Expressions}

The invariants of stress and strain and their increments suffice to express the constitutive relations for isotropic materials. The stress invariants are defined as

\begin{tabular}{p{0.5in}p{3.4in}p{0.5in}}
 & $p \equiv \boldsymbol{\sigma }:\boldsymbol{I}/3$ & (26 bis) \\
 & $q={\left({{3}\boldsymbol{q}\boldsymbol{:}\boldsymbol{q}}/{2}\right)}^{\frac{1}{2}}$ & (64) \\
\end{tabular}

\noindent
where $\boldsymbol{q}$ denotes the deviator stress tensor:

\begin{tabular}{p{0.5in}p{3.4in}p{0.5in}}
 & $\boldsymbol{q}\ \equiv \boldsymbol{\sigma }-p\boldsymbol{I}$ & (65) \\
\end{tabular}

\noindent
The strain increment invariants are work-conjugate to the stress invariants:

\begin{tabular}{p{0.5in}p{3.4in}p{0.5in}}
 & $\delta \epsilon \equiv \boldsymbol{I}:\delta \boldsymbol{\epsilon }$ & (28 bis) \\
 & $\delta \gamma ={\left({2\delta \boldsymbol{\gamma }\boldsymbol{:}\delta \boldsymbol{\gamma }}/{3}\right)}^{\frac{1}{2}}$ & (66) \\
\end{tabular}

\noindent
where $\boldsymbol{\gamma }$ denotes the deviator-strain tensor:

\begin{tabular}{p{0.5in}p{3.4in}p{0.5in}}
 & $\delta \boldsymbol{\gamma }\ \equiv \delta \boldsymbol{\epsilon }- (\boldsymbol{I}/3)\ \delta \epsilon$ & (67) \\
\end{tabular}

\noindent
The macroscopic relations are

\begin{tabular}{p{0.5in}p{3.4in}p{0.5in}}
 & $\delta p=K\delta \epsilon +J\delta \gamma $ & (68) \\
 & $\delta q=J\delta \epsilon +3G\delta \gamma $ & (69) \\
\end{tabular}

\noindent
where $K,\ J,$ and $G$ are respectively the macroscopic bulk, cross and shear moduli.

These moduli consist of uniform and differential components. From Eq. (25):

\begin{tabular}{p{0.5in}p{3.4in}p{0.5in}}
 & $K=K_u-{\left({\omega }^2_pK^2_u+2{\omega }_p{\omega }_qJ_uK_u+{\omega }^2_qJ^2_u\right)}\ /\ {S_c}$ & (70) \\
 & $J=J_u-{\left[{\omega }^2_pJ_uK_u+\ {\omega }_p{\omega }_q\left(J^2_u+3K_uG_u\right)+3{\omega }^2_qJ_uG_u\right]}\ /\ {S_c}$ & (71) \\
 & $3G=3G_u-{\left({\omega }^2_pJ^2_u+6{\omega }_p{\omega }_qJ_uG_u+9{\omega }^2_qG^2_u\right)}\ /\ {S_c}$ & (72) \\
\end{tabular}

\noindent
where $K_u,\ J_u,\ {and\ G}_u$${}^{\ }$denote respectively the uniform bulk, cross and shear moduli; ${\omega }_p,$ ${\omega }_q$ denote the invariants of the normalized coupling tensor; and $S_c$ denotes the composite contraction-swelling modulus:

\begin{tabular}{p{0.5in}p{3.4in}p{0.5in}}
 & $S_c=S+\ {\omega }^2_pK_u+2{\omega }_p{\omega }_{q\ }J_u+3{\omega }^2_qG_u$ & (73) \\
\end{tabular}

\noindent
This composite modulus augments the contraction-swelling modulus with the element's uniform stiffness.

\subsection{Linear Solution}

The simplest internal energy potential that conserves energy for a linear isotropic material with a variable specific volume is a quadratic function in volumetric and deviatoric strain invariants and an inverse function of specific volume

\begin{tabular}{p{0.5in}p{3.4in}p{0.5in}}
 & $E\left(\epsilon ,\gamma ,\nu \right)=p_r(k{\epsilon }^2+3g{\gamma }^2)/2\nu $ & (74) \\
\end{tabular}

\noindent
where $p_r$ denotes an arbitrarily selected reference pressure; $k$ and $g\ $denote the  non-dimensional bulk and shear indices corresponding to $p_r$. The  complementary internal energy potential is quadratic in the stress invariants and proportional to specific volume

\begin{tabular}{p{0.5in}p{3.4in}p{0.5in}}
 & $C(p,q,\nu )=\nu (p^2/k+q^2/3g)/2p_r$ & (75) \\
\end{tabular}

\noindent
The strain and stress invariants are linearly related. From Eqs. (40) and (44):

\begin{tabular}{p{0.5in}p{3.4in}p{0.5in}}
 & $\epsilon \equiv \frac{\partial C}{\partial p}=\left(\frac{1}{kp_r}\right)\nu p$ & (76) \\
 & $\gamma \equiv \frac{\partial C}{\partial q}=\left(\frac{1}{3gp_r}\right)\nu q$ & (77) \\
 & $p\ \equiv \frac{\partial E}{\partial \epsilon }=\left(\frac{kp_r}{\nu }\right)\epsilon $ & (78) \\
 & $q\ \equiv \frac{\partial E}{\partial \gamma }=\left(\frac{3gp_r}{\nu }\right)\gamma $ & (79) \\
\end{tabular}

\noindent
The uniform moduli are partial derivatives of the stress invariants with respect to strain:

\begin{tabular}{p{0.5in}p{3.4in}p{0.5in}}
 & $K_u\ \equiv \frac{{\partial }^2E}{\partial {\epsilon }^2}=kp_r/\nu $ & (80) \\
 & $J_u\ \equiv \frac{{\partial }^2E}{\partial \epsilon \partial \gamma }=0$ & (81) \\
 & $G_u\ \equiv \frac{1}{3}\frac{{\partial }^2E}{{\partial }^2\gamma }=gp_r\mathrm{/}\mathrm{\nuup }$ & (82) \\
\end{tabular}

\noindent
These moduli are the linearly scaled versions of their mesoscopic counterparts (that is, the particle bulk and shear moduli ($kp_r,\ gp_r$) [51]).

The coupling tensor coefficients are cross derivatives of the complementary internal energy potential. From Eqs. (50) and (75):

\begin{tabular}{p{0.5in}p{3.4in}p{0.5in}}
 & ${\mathit{\Omega}}_p\ \equiv \frac{{\partial }^2C}{\partial \nu \partial p}=\frac{p}{kp_r}$ & (83) \\
 & ${\mathit{\Omega}}_q\ \equiv \frac{{\partial }^2C}{\partial \nu \partial q}=\frac{q}{3gp_r}$ & (84) \\
\end{tabular}

\noindent
The normalized coefficients are

\begin{tabular}{p{0.5in}p{3.4in}p{0.5in}}
 & ${\omega }_p={\left[1+{\left(\frac{k}{3g}\right)}^2{\left(\frac{q}{p}\right)}^2\right]}^{-\frac{1}{2}}$ & (85) \\
 & ${\omega }_q=\left(\frac{k}{3g}\right)\left(\frac{q}{p}\right){\omega }_p$ & (86) \\
\end{tabular}

\noindent
The normalized coupling tensor introduces shear stress effects to the uniform moduli. Substituting into Eqs. (70) through (72) yields

\begin{tabular}{p{0.5in}p{3.4in}p{0.5in}}
 & $K=\frac{kp_r}{\nu }\frac{T+\frac{\left(\frac{k}{3g}\right){\left(\frac{q}{p}\right)}^2}{\left[1+{\left(\frac{k}{3g}\right)}^2{\left(\frac{q}{p}\right)}^2\right]}}{T_c}$ & (87) \\
 & $J=-\frac{kp_r}{\nu }\frac{\frac{\left(\frac{q}{p}\right)}{\left[1+{\left(\frac{k}{3g}\right)}^2{\left(\frac{q}{p}\right)}^2\right]}}{T_c}$ & (88) \\
 & $G=\frac{gp_r}{\nu }\frac{T+\frac{1}{\left[1+{\left(\frac{k}{3g}\right)}^2{\left(\frac{q}{p}\right)}^2\right]\ }}{T_c}$ & (89) \\
\end{tabular}

\noindent
where $T$ and $T_c$ denote the relative-contraction-swelling and composite relative-contraction-swelling moduli respectively:

\begin{tabular}{p{0.5in}p{3.4in}p{0.5in}}
 & $T\ \equiv \left(\frac{S}{K_{u|q=0\ }}\right)\ ={S}/\left({kp_r}/{\nu }\right)$ & (90) \\
 & $T_c={S_c}/\left({kp_r}/{\nu }\right)=T+{\left[1+\left(\frac{k}{3g}\right){\left(\frac{q}{p}\right)}^2\right]}{\left[1+{\left(\frac{k}{3g}\right)}^2{\left(\frac{q}{p}\right)}^2\right]}^{-1}$ & (91) \\
\end{tabular}

\noindent
Although the macroscopic moduli are inversely proportional to specific volume, the linear scaling evident in the uniform moduli is absent in the macroscopic moduli. Eqs. (87) through (89) extend the particle stress model [51] across state space without imposing perfectly linear scaling.

The macroscopic moduli assume simpler forms at isotropic loading states. From Eqs. (87) through (89),

\begin{tabular}{p{0.5in}p{2.65in}p{1.25in}}
 & $K=\frac{\frac{kp_r}{\nu }}{\frac{1}{T}+1}=\frac{S}{1+T}$ & $for\ q=0$ (92) \\
 & $J=0$ & $for\ q=0$ (93) \\
 & $G=G_u=\frac{gp_r}{\nu }$ & $for\ q=0$ (94) \\
\end{tabular}

\noindent
As the bulk index increases ($T\ \to 0$), the macroscopic bulk modulus approaches the contraction-swelling modulus ($S$).

The packing pressure follows directly from the complementary internal energy potential:

\begin{tabular}{p{0.5in}p{3.4in}p{0.5in}}
 & $\phi = \frac{\partial C}{\partial \nu }={\left[p^2+\left(\frac{k}{3g}\right)q^2\right]}\ /\ {2kp_r}$ & (95) \\
\end{tabular}

\noindent
Differentiating (95) yields

\begin{tabular}{p{0.5in}p{3.4in}p{0.5in}}
 & ${\delta \phi }/{\phi }={2\left[p\delta p+\left(\frac{k}{3g}\right)q\delta q\right]} {\left[p^2+\left(\frac{k}{3g}\right)q^2\right]}^{-1}$ & (96) \\
\end{tabular}

\noindent
At multi-scale equilibrium, specific volume and packing pressure are constant. Integrating Eq. (96) for mesoscopic equilibrium ($\delta \phi = \delta \nu = 0$) yields

\begin{tabular}{p{0.5in}p{3.4in}p{0.5in}}
 & $b\left(p,q,\nu \right)=p^2+\left(\frac{k}{3g}\right)q^2-p^2_o=\ 0$ & (97) \\
\end{tabular}

\noindent
The integration constant ($p_o$) is a function of specific volume alone.

Expressions for the contraction-swelling modulus complete this solution. As noted in sub-section 3.2, the uniform bulk modulus can be estimated from the shear modulus if the Poisson's ratio ($\rho $) for the family of solid phases is constant:

\begin{tabular}{p{0.5in}p{2.65in}p{1.25in}}
 & $k=2g(1+\rho )/3(1-2\rho )\ $ &      $for\ q=0$ (98) \\
\end{tabular}

\noindent
The lower limit on the contraction-swelling modulus follows from Eq. (32):

\begin{tabular}{p{0.5in}p{2.65in}p{1.25in}}
 & $S \geq {\left(\frac{\delta \epsilon }{\delta p_o}\right)}^{-1}$ &       $for\ q=0$ (99) \\
\end{tabular}

\noindent
For a linear relation in $\nu -ln (p_o/p_r) $ space

\begin{tabular}{p{0.5in}p{3.4in}p{0.5in}}
 & $S = \nu p/\kappa $ & (100) \\
\end{tabular}

\noindent
where $\kappa $ denotes the contraction-swelling index in $\nu -ln (p_o/p_r) $ space:

\begin{tabular}{p{0.5in}p{2.65in}p{1.25in}}
 & $\delta \nu =\ -\ \kappa \delta p/p$ & $for\ q=0$ (101) \\
\end{tabular}

\noindent
For a linear relation in $ln \nu -ln (p_o/p_r) $ space

\begin{tabular}{p{0.5in}p{2.65in}p{1.25in}}
 & $S = p/{\kappa }^*$ & $for\ q=0$ (102) \\
\end{tabular}

\noindent
where ${\kappa }^*$ denotes the modified contraction-swelling index in $ln \nu -ln (p_o/p_r) $ space:

\begin{tabular}{p{0.5in}p{2.65in}p{1.25in}}
 & $\delta \nu /\nu =\ -\ {\kappa }^*\ \delta p/p$ & $for\ q=0$ (103) \\
\end{tabular}

\noindent
The value of the contraction-swelling modulus is determined at isotropic loading and extrapolated to non-isotropic loading states.

\subsection{Non-Linear Solution}

The non-linear specialization for isotropic materials is based on the generally accepted expression for the small-strain shear modulus of a range of silts and clays [40]:

\begin{tabular}{p{0.5in}p{2.65in}p{1.25in}}
 & $G_u=gp_r{\left(\frac{p}{p_r}\right)}^n{\nu }^{-a}$ & $for\ q=0$ (104) \\
\end{tabular}

\noindent
where $n$ and $a$ are non-dimensional and denote respectively the element's shape and scaling coefficients. (Typical data: $15,000<gp_r<25,000,\ a\approx 2.4$ and [52,53] $n=0.50$ for smooth spherical contacts, $n=0.33\ $for angular contacts.) The corresponding uniform bulk modulus is

\begin{tabular}{p{0.5in}p{2.65in}p{1.25in}}
 & $K_u=kp_r{\left(\frac{p}{p_r}\right)}^n{\nu }^{-a}$ & $for\ q=0$ (105) \\
\end{tabular}

\noindent
where the bulk index can be determined from Eq. (98) assuming a constant Poisson's ratio for the family of solid phases.

The internal energy potential corresponding to both expressions is

\begin{tabular}{p{0.5in}p{3.4in}p{0.5in}}
 & $E\left(\epsilon ,\gamma ,\nu \right)={p_r{\left[k\left(1-n\right)\psi {\nu }^{-a}\right]}^{\frac{2-n}{1-n}}{\nu }^a}/{k\left(2-n\right)}$ & (106) \\
\end{tabular}

\noindent
where $\psi$ denotes an equivalent volumetric strain defined by

\begin{tabular}{p{0.5in}p{3.4in}p{0.5in}}
 & $\psi^2\ \equiv {\epsilon }^2+{\gamma }^2/h$ & (107) \\
\end{tabular}

\noindent
and where $h$ denotes the effective uniform stiffness ratio defined by

\begin{tabular}{p{0.5in}p{3.4in}p{0.5in}}
 & $h\ \equiv \ k(1-n)/3g$ & (108) \\
\end{tabular}

\noindent
The corresponding complementary strain energy potential is

\begin{tabular}{p{0.5in}p{3.4in}p{0.5in}}
 & $C\left(p,q,\nu \right)={p_rr^{2-n}{\nu }^a}\ /\ {k\left(2-n\right)\left(1-n\right)}$ & (109) \\
\end{tabular}

\noindent
where $r$ denotes the equivalent pressure ratio defined by

\begin{tabular}{p{0.5in}p{3.4in}p{0.5in}}
 & $r^2\ \equiv (p^2+hq^2)/p^2_r$ & (110) \\
\end{tabular}

\noindent
The strain and stress invariants are power functions of the equivalent pressure ratio and the specific volume. From Eqs. (40) and (44):

\begin{tabular}{p{0.5in}p{3.4in}p{0.5in}}
 & $\epsilon =\left(\frac{1}{k\left(1-n\right)p_rr^n}\right){\nu }^ap$ & (111) \\
 & $\gamma =\left(\frac{1}{3gp_rr^n}\right){\nu }^aq$ & (112) \\
 & $p=p_r{\left(\frac{k\left(1-n\right)^n}{{\nu }^a}\right)}^{\frac{1}{1-n}}\epsilon $ & (113) \\
 & $q=p_r{\left(\frac{k\left(1-n\right)^n}{{\nu }^a}\right)}^{\frac{1}{1-n}}\left(\frac{1}{h}\right)\gamma $ & (114) \\
\end{tabular}

\noindent
The uniform moduli are given by

\begin{tabular}{p{0.5in}p{3.4in}p{0.5in}}
 & $K_u=\frac{kp_rr^n}{{\nu }^a}\left\{1-{nh{\left(\frac{q}{p}\right)}^2}/{\left[1+h{\left(\frac{q}{p}\right)}^2\right]}\right\}$ & (115) \\
 & $J_u=\frac{kp_rr^n}{{\nu }^a}\left\{{n\left(\frac{q}{p}\right)}/{\left[1+h{\left(\frac{q}{p}\right)}^2\right]}\right\}$ & (116) \\
 & $G_u=\frac{kp_rr^n}{3h{\nu }^a}\left\{1-{n}/{\left[1+h{\left(\frac{q}{p}\right)}^2\right]}\right\}$ & (117) \\
\end{tabular}

\noindent
These moduli are scaled versions of their mesoscopic counterparts ($\nu = 1$).

The coupling tensor coefficients follow directly from the complementary internal energy potential. From Eqs. (50) and (109):

\begin{tabular}{p{0.5in}p{3.4in}p{0.5in}}
 & ${\mathit{\Omega}}_p={a{\nu }^{a-1}p}/{kp_r\left(1-n\right)r^n}$ & (118) \\
 & ${\mathit{\Omega}}_q={a{\nu }^{a-1}q}/{3gp_rr^n}$ & (119) \\
\end{tabular}

\noindent
The normalized coefficients are

\begin{tabular}{p{0.5in}p{3.4in}p{0.5in}}
 & ${\omega }_p={\left[1+h^2{\left(\frac{q}{p}\right)}^2\right]}^{-\frac{1}{2}}$ & (120) \\
 & ${\omega }_q=h\left(\frac{q}{p}\right){\omega }_p$ & (121) \\
\end{tabular}

\noindent
The macroscopic moduli are given by

\begin{tabular}{p{0.5in}p{3.4in}p{0.5in}}
 & $K=\frac{kp_rr^n}{{\nu }^a}{\left\{T\left[1-\frac{nh{\left(\frac{q}{p}\right)}^2}{1+h{\left(\frac{q}{p}\right)}^2}\right]+\frac{\left(1-n\right)h{\left(\frac{q}{p}\right)}^2}{1+h^2{\left(\frac{q}{p}\right)}^2}\right\}}\ /\ {T_c}$ & (122) \\
 & $J=\frac{kp_rr^n}{{\nu }^a}{\left\{T\left[\frac{n\left(\frac{q}{p}\right)}{1+h{\left(\frac{q}{p}\right)}^2}\right]-\frac{\left(1-n\right)\left(\frac{q}{p}\right)}{1+h^2{\left(\frac{q}{p}\right)}^2}\right\}}\ /\ {T_c}$ & (123) \\
 & $G=\frac{kp_rr^n}{3h{\nu }^a}{\left\{T\left[1-\frac{n}{1+h{\left(\frac{q}{p}\right)}^2}\right]+\frac{1-n}{1+h^2{\left(\frac{q}{p}\right)}^2}\ \right\}}\ /\ {T_c}$ & (124) \\
\end{tabular}

\noindent
where

\begin{tabular}{p{0.5in}p{3.4in}p{0.5in}}
 & $T_c=T+{1+h{\left(\frac{q}{p}\right)}^2}\ /\ {1+h^2{\left(\frac{q}{p}\right)}^2}$ & (125) \\
 & $T={S{\nu }^a}/{kp_rr^n\ }$ & (126) \\
\end{tabular}

\noindent
These moduli take simpler forms at isotropic loading states. From Eqs. (122) through (126):

\begin{tabular}{p{0.5in}p{2.65in}p{1.25in}}
 & $K=\frac{kp_r{\left(\frac{p}{p_r}\right)}^n{\nu }^{-a}}{T^{-1}+1}={S}\ /\ {1+T}$ & $for\ q=0$ (127) \\
 & $J=0$ & $for\ q=0$ (128) \\
 & $G=G_u={gp_r{\left(\frac{p}{p_r}\right)}^n}/{{\nu }^a}$ & $for\ q=0$ (129) \\
\end{tabular}

\noindent
As the family of solid phases approaches volumetric incompressibility ($\rho \to \frac{1}{2}$) the value of the macroscopic bulk modulus reaches a linear function of pressure:

\begin{tabular}{p{0.5in}p{2.65in}p{1.25in}}
 & $K=S$ & $for\ q=0$ (130) \\
\end{tabular}

\noindent
while the shear modulus remains a proper fractional power of pressure. The expressions for the contraction-swelling modulus are the same as those for the linear solution (given by Eqs. (100) and (102)).

The packing pressure follows directly from the complementary energy potential:

\begin{tabular}{p{0.5in}p{3.4in}p{0.5in}}
 & $\phi =\frac{\partial C}{\partial \nu }={ap_rr^{2-n}{\nu }^{a-1}}/{k(2-n)(1-n)}$ & (131) \\
\end{tabular}

\noindent
Differentiating (131) gives

\begin{tabular}{p{0.5in}p{3.4in}p{0.5in}}
 & ${\delta \phi }/{\phi }=\left[{(a-1)\delta \nu }/{\nu }\right]+\left[{\left(2-n\right)\delta r}/{r}\right]$ & (132) \\
\end{tabular}

\noindent
Integrating Eq. (132) for multi-scale equilibrium at constant packing pressure ($\delta \phi = \delta \nu = 0$) yields

\begin{tabular}{p{0.5in}p{3.4in}p{0.5in}}
 & $b\left(p,q,\nu \right)=p^2+hq^2-p^2_o=\ 0$ & (133) \\
\end{tabular}

\noindent
The integration constant ($p_o$) changes with specific volume only.

\begin{figure}[ht!]
\centering
\includegraphics[scale=0.5]{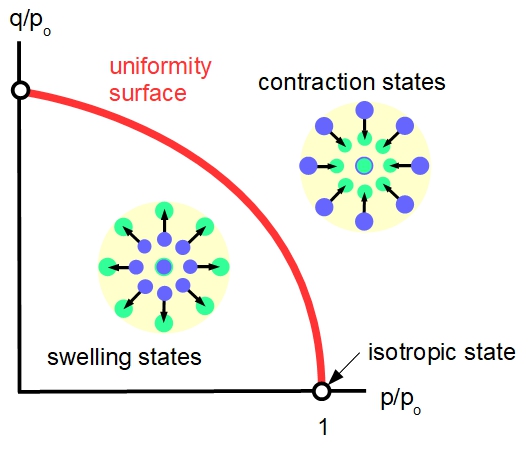}
\caption{Uniformity Surface for an Isotropic Material}
\label{fig:Fig8}
\end{figure}

Figure \ref{fig:Fig8} illustrates the uniformity surface described by Eq. (133) in stress space normalized with respect to isotropic loading pressure ($p_o$). This surface partitions normalized stress space into contraction and swelling sub-domains. Only stress increments directed along this surface produce purely uniform straining.

\section{Discussion}

The decomposition of deformation into uniform and differential components in the mesoscopic model allows distinct representations at different scales. The macroscopic compliance and elasticity tensors consist of uniform and differential components. Their uniform properties map from macroscopic to mesoscopic scales. Their mapping accounts for the absence of interstitial content at mesoscopic scale by factoring the high-level observed properties by a power function of specific volume. In other words, the uniform component of the higher level observed properties is just an upscale version of multi-constancy properties that are defined at mesoscopic scale. This uniform component can be measured directly by applying stress increments that do not alter the specific volume. The differential properties of these macroscopic compliance and elasticity tensors  are stress space constraints based on property ratios and stress ratios, but not values themselves. The gradient to the uniformity surface can be measured by applying stress increments that alter specific volume. In crossing scales, the uniformity surface reduces to a point on the contraction-swelling curve. Although the surface has no scaled mesoscopic counterpart, the macroscopic contraction-swelling modulus is related to the tangential slope of the curve of the contraction-swelling curve. That is, higher level observed changes in the constraint surface map to mesoscopic changes at multi-scale equilibrium.

\begin{figure}[ht!]
\centering
\includegraphics[scale=0.5]{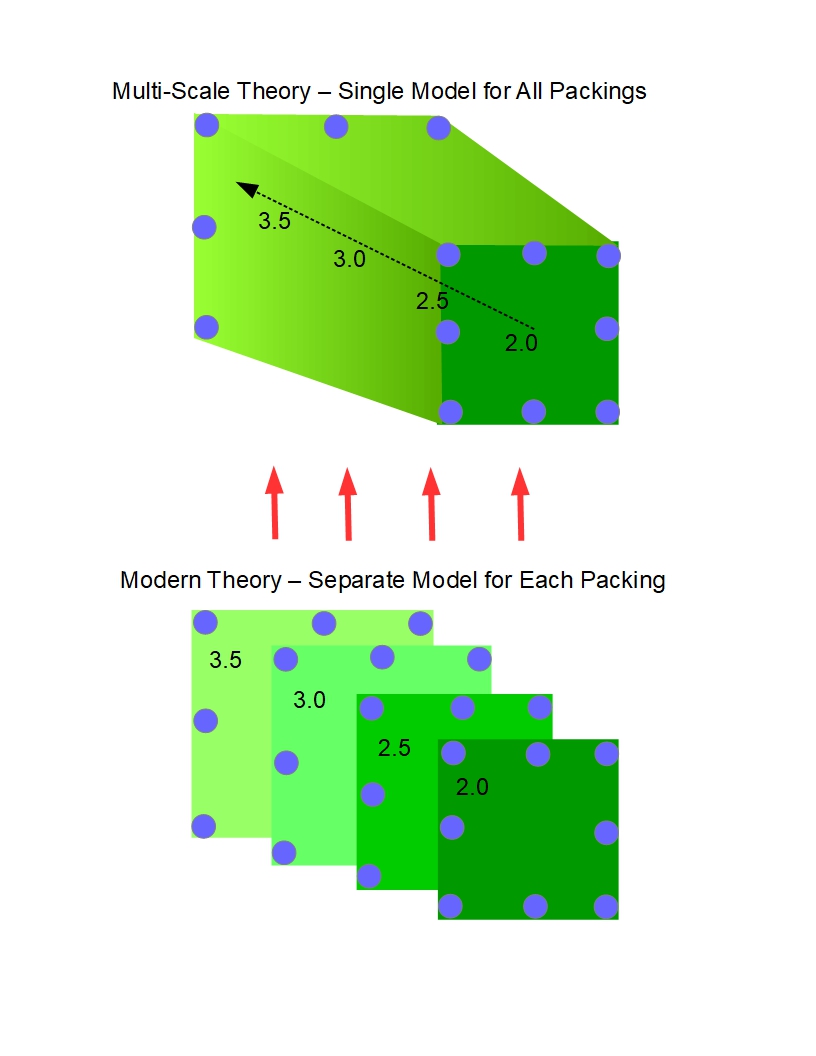}
\caption{Multi-Scale and Modern Theories of Elasticity}
\label{fig:Fig9}
\end{figure}

This multi-scale theory includes the modern theory of elasticity as a special case. That is the case in which the contraction-swelling modulus is indefinitely large. From this perspective, the modern theory's scope is limited to materials that exhibit negligible contraction and swelling; that is, to materials that can be modeled by a single solid phase of constant specific volume. In other words, the present theory extends the modern theory from one for a prescribed specific volume to one that models continuous variation across a family of solid phases of distinct specific volumes. Figure \ref{fig:Fig9} depicts the relation between the continuous solid phase represented by this multi-scale theory and the family of discrete solid phases each represented as a different material by the predecessor modern theory.

Table \ref{T1} lists the expressions for the bulk and shear moduli of isotropic materials \textit{at isotropic loading states} ($q=0$). The upper three rows apply to soils and aggregates, while the lower two rows apply to porous solids. The exponents of the power functions of pressure for the bulk and shear moduli are identical only for ideal porous solids. The moduli diverge moving up the Table. The ideal aggregate classes exhibit a bulk modulus linear in pressure, and possibly specific volume, and a shear modulus that is indefinitely large.

\subsection{Data for Soils and Aggregates}

Soils and tire-derived aggregates (TDA) belong to the three aggregate classes listed in Table 1. In all three classes, differential volumetric straining dominates volumetric straining. As a first approximation, volumetric straining represents changes in particle proximity alone.

\begin{table}[h]
\centering
\begin{tabular}{|M{2.9in}|M{1.1in}|M{1.0in}|M{1.1in}|N} \hline
\multirow{2}{1.0in}{Class}  & \multicolumn{2}{P{2.1in}|}{Bulk Modulus} & \multirow{2}{1.0in}{Shear Modulus} \\
 & \textit{S = }$\frac{\nu p_o}{\kappa }$\textit{} & \textit{S = }$\frac{p_o}{{\kappa }^*}$ &  \\[5pt] \hline
Ideal Aggregates & ${\frac{\nu p_o}{\kappa }}$ & ${\frac{p_o}{{\kappa }^*}}$ & ${\mathrm{\infty}}$ \\[5pt] \hline
Ideal Distortionally Compliant Aggregates  & ${\frac{\nu p_o}{\kappa }}$ & ${\frac{p_o}{{\kappa }^*}}$ & ${\frac{gp_r}{{\nu }^a}{\left(\frac{p_o}{p_r}\right)}^n}$ \\[8pt] \hline
Aggregates  & ${\frac{\frac{\nu p_o}{\kappa }}{1+\nu \left(\frac{{\nu }^a}{\kappa k}\right){\left(\frac{p_o}{p_r}\right)}^{\left(1-n\right)}}}$ & ${\frac{\frac{p_o}{{\kappa }^*}}{1+{\left(\frac{{\nu }^a}{{\kappa }^*k}\right)\left(\frac{p_o}{p_r}\right)}^{\left(1-n\right)}}}$ &
${\frac{gp_r}{{\nu }^a}{\left(\frac{p_o}{p_r}\right)}^n}$ \\[11pt] \hline
Porous Solids  & ${\frac{\frac{kp_r}{{\nu }^a}{\left(\frac{p_o}{p_r}\right)}^n}{1+\left(\frac{1}{\nu }\right)\left(\frac{\kappa k}{{\nu }^a}\right){\left(\frac{p_o}{p_r}\right)}^{n-1}}}$ & ${\frac{\frac{kp_r}{{\nu }^a}{\left(\frac{p_o}{p_r}\right)}^n}{1+\left(\frac{{\kappa }^*k}{{\nu }^a}\right){\left(\frac{p_o}{p_r}\right)}^{n-1}}}$ &
${\frac{gp_r}{{\nu }^a}{\left(\frac{p_o}{p_r}\right)}^n}$ \\[13pt] \hline
Ideal Porous Solids & ${\frac{kp_r}{{\nu }^a}{\left(\frac{p_o}{p_r}\right)}^n}$ & ${\frac{kp_r}{{\nu }^a}{\left(\frac{p_o}{p_r}\right)}^n}$ &
${\frac{gp_r}{{\nu }^a}{\left(\frac{p_o}{p_r}\right)}^n}$ \\[13pt] \hline
\end{tabular}
\caption{Bulk and Shear Moduli at Isotropic Loading States}
\label{T1}
\end{table}

Figures~\ref{fig:Fig10} and \ref{fig:Fig11} compare predictions using Eqs. (29) with (102) and the non-linear solution to published data for fine Ottawa sand under isotropic unloading [54]. The contraction-swelling index selected to fit the data for \textit{both loose and dense} samples is ${\kappa }^*=0.0001.$ A shape factor of $n=0.5$ models the particles as relatively rounded. The best fits for loose ($\nu =1.836$) and dense ($\nu =1.665$) data are $\frac{k}{{\nu }^a}=700$ and $\frac{k}{{\nu }^a}=1000$ respectively, given a reference pressure of 1 atmosphere ($p_r=1$). These values yield $a=3.648$ and $k=6423$ for both packings. The permanent volumetric strains that align the curves based on these coefficients are listed in the Figure legends. The upper limit on each curve is the pressure at the end of reloading and the onset of unloading. Combining these volumetric properties with the small-strain shear data for Ottawa sand [55] gives a shear index of $g=4783$ (for $a=3.648$ and $p_r\ =1$). This shear index combined with the derived bulk index gives a uniform Poisson's ratio of $\rho =0.2$. The correspondence between the theory and this independently published data for loose and dense packing and for volumetric and distortional straining is quite encouraging.

%
%
%

Tire-derived aggregate is twenty times more compressible than sand. Its compressibility is due to the presence of tire chips, which are a common additive to municipal solid waste. TDAs exhibit visible amounts of uniform volumetric deformation. Figure~\ref{fig:Fig12} compares the pore volume changes to particle compression under isotropic reloading and unloading [56]. The solid line identifies the volumetric strain. The dashed line identifies the differential strain of the interstitial phase accumulated during reloading. The volumetric strain of the aggregate is still predominantly differential.

%

\subsection{Data for Porous Solids}

Consolidated rock, concrete and ceramics belong to the porous solid classes listed in Table 1. Uniform volumetric straining dominates in these two classes. The ideal porous solid class is the theory's regular limit: its contraction-swelling modulus is large enough to disregard any differential volumetric straining ($S\to \infty $). As a first approximation, volumetric straining is straining of a solid phase with constant specific volume. The bulk and shear moduli are functions of porosity, increasing as porosity decreases.

Data for porous solids is typically presented in normalized form with respect to the modulus of the mesoscopic solid constituent (its projected value at zero porosity). Anderson [57] studied the dependence of bulk and shear moduli on volume per ion pair for a wide variety of materials. Anderson's relation in normalized form [58] is

\begin{tabular}{p{0.5in}p{2.65in}p{1.25in}}
 & $\frac{K}{K_{\left(u\mathrel{\left|\vphantom{u \nu =1}\right.\kern-\nulldelimiterspace}\nu =1\right)}}\approx \frac{K_u}{K_{\left(u\mathrel{\left|\vphantom{u \nu =1}\right.\kern-\nulldelimiterspace}\nu =1\right)}}=\frac{G_u}{G_{\left(u\mathrel{\left|\vphantom{u \nu =1}\right.\kern-\nulldelimiterspace}\nu =1\right)}}={\left(1-\eta \right)}^a$ & $T\to \infty $ (134) \\
\end{tabular}

\noindent
$K_{u|\nu =1}$ denotes the uniform bulk modulus (of the constituent material) evaluated at zero porosity. Relation (134) is identical to Eqs. (115) and (117) at isotropic loading. Anderson proposed $a=5$ for oxide ceramics in general, and $K_{u|\nu =1}=252GPa$ for alumina specifically. Munro [58] developed this relation for the bulk modulus of high-purity alumina using effective medium theory and found $K_{u|\nu =1}=252GPa,$ $a=2.1$ to be optimal, based on 16 references to empirical data.

Krief \textit{et al.} [59] used experimental data and Pickett's [60] empirical result (which assumes a system Poisson's ratio approximately equal to the mineral ratio) to show that for dry rock, $\left(1-\beta \right)={\left(1-\eta \right)}^{m\left(\eta \right)}$, where $\beta $ is Biot's first coefficient and $m\left(\eta \right)=\frac{3}{1-\eta }$. Knackstedt \textit{et al.} [61] relied on this to write

\begin{tabular}{p{0.5in}p{2.65in}p{1.25in}}
 & $\frac{K}{K_{u|\nu =1}}\approx \frac{K_u}{K_{u|\nu =1}}=\frac{G_u}{G_{u|\nu =1}}={(1-\eta )}^{\left(\frac{3}{1-\eta }\right)}$ & $T\to \infty $ (135) \\
\end{tabular}

\noindent
Knackstedt \textit{et al. }reported finite element simulation results for the elastic properties of dry cemented sandstone that support a non-linear relation between these moduli and porosity ($a\napprox 1$) and show an accurate reproduction of the Krief \textit{et al.} empirical relation between shear modulus and porosity.

\subsection{Critical State Soil Mechanics}

Critical State Soil Mechanics (CSSM) [5,33] describes soils in the ideal aggregate class. It models a soil sample as `a random aggregate of irregular solid particles of diverse sizes which tear, rub, scratch, chip and even bounce against each other during the process of continuous deformation' and applies at the length scale at which flow and deformation appear continuous.

The CSSM macroscopic bulk modulus is linearly related to effective pressure. The model's internal energy potential assumes a distortionally rigid solid phase ($g\to \infty $) [62]. The volumetric rigidity of the solid phase ($k\to \infty $) necessarily follows from this assumption (Eq. (98)). In terms of the present theory, volumetric straining is purely differential and volumetric intra-particle straining is negligible. The particles themselves do not store strain energy.

\subsubsection{Isotropic Loading States}

Improvements to the CSSM model are possible based on the present theory. The simplest linear solution that introduces distortional strain energy to the CSSM potential, while ignoring both shape and scaling properties ($n=0,\ a=0$). The solid phase is then volumetrically compliant ($0\le \rho <\frac{1}{2},k\le \infty , 0\le T < \infty $). The corresponding expressions for the bulk, cross and shear moduli are

\begin{tabular}{p{0.5in}p{3.4in}p{0.5in}}
 & $K={\left[\left({1/kp_r }\right) + \kappa / \nu p \right]}^{-1}$ & (136) \\
 & $J=0$ & (137) \\
 & $G=gp_r$ & (138) \\
\end{tabular}

\noindent
\textit{Three} coefficients describe the properties of a soil sample: its contraction-swelling index ($\kappa $ or ${\kappa }^* $) and the bulk and shear indices of its solid phase ($k$ and $g$).

Zytynski \textit{et al.} [2] noted that selecting a constant shear modulus may lead to negative-valued Poisson's ratios, which is physically unrealistic for soils. Eqs (136) through (138) facilitate an energetically conservative model through  an independent selection of a constant shear modulus and a positive-valued Poission's ratio for the family of solid phases. Letting this Poisson's ratio increase to $\rho = \frac{1}{2}$ recovers the CSSM model in two coefficients and energetically conservative form, with a constant shear modulus.

A further enhancement involves coupling the macroscopic properties to specific volume. Adding linear scaling ($a=1$) couples the strain energy contribution to specific volume and predicts a shear modulus that is inversely proportional to specific volume [51]. The corresponding expressions for the macroscopic moduli are Eqs. (92), (93), (94) and (100). Adding non-linear shape and scaling ($n>0,\ a>1$) yields the commonly accepted expression for small-strain shear modulus [40]. The corresponding expressions are Eqs. (127), (128), (129) and (100). This is the class in the third topmost row of Table \ref{T1}.

Letting the solid phase's Poisson's ratio increase to $\rho = \frac{1}{2}$ recovers the CSSM model in two coefficients, but now with the commonly accepted expression for small-strain shear modulus. This is the ideal distortionally complaint class in the second topmost row of Table \ref{T1}.

\subsubsection{Non-Isotropic Loading States}

The bulk, cross and shear moduli for non-isotropic loading states extrapolate on the CSSM model. For the simplest enhancement in sub-section 6.3.1, which introduces distortional compliance, but ignores both shape and scaling properties ($n=0,\ a=0$), the present theory yields

\begin{tabular}{p{0.5in}p{3.4in}p{0.5in}}
 & $K=kp_r\ \frac{W+h{\left(\frac{q}{p}\right)}^2}{W+1+h{\left(\frac{q}{p}\right)}^2\ }$ & (139) \\
 & $J=-\ kp_r\frac{\frac{q}{p}}{W+1+h{\left(\frac{q}{p}\right)}^2}$ & (140) \\
 & $G=gp_r\frac{W+1}{W+1+h{\left(\frac{q}{p}\right)}^2}$ & (141) \\
\end{tabular}

\noindent
where

\begin{tabular}{p{0.5in}p{3.4in}p{0.5in}}
 & $W=\frac{\nu p}{\kappa kp_r}\left[1+h^2{\left(\frac{q}{p}\right)}^2\right]$ & (142) \\
\end{tabular}

\noindent
These expressions are valid extensions of the CSSM model provided that the solid phase is not perfectly incompressible ($\rho < \frac{1}{2}$).

The present theory does not encompass the special case of a family of incompressible solid phases with different specific volumes at non-isotropic loading states. That is, the CSSM model cannot be recovered from the present theory at these states. Letting the solid phase's bulk index increase without limit at any state of non-zero shear stress in either the linear solution or the non-linear solution (Eqs. (87) through (91) or Eqs. (122) through (124) respectively) leads to a singularity.

\subsection{Future Considerations}

The singularities that arise at non-isotropic loading states for volumetrically incompressible solid phases expose a limitation of the present theory. Its inability to recover the original CSSM model, which assumes volumetric incompressibility, constrains the present theory's scope to those materials with solid phases that exhibit some volumetric compliance; that is, phases with a uniform Poisson's ratio in the range $0 \le \rho <\frac{1}{2}$.

Broadening the theory's scope to include perfect volumetric incompressibility ($\rho = \frac{1}{2}$) calls for an intertheoretic solution [64]. The present theory and the CSSM model achieve different objectives. The CSSM model [63], like the bulk solids' model of Jenike and Shield [65], identifies stable loci of states of continuous plastic flow (critical states). Both models relate the shear stress to effective pressure at any critical state through a frictional constant, $q=Mp'$. On the other hand, the present theory resolves the energy conservation issue at states away from continuous plastic flow states considering friction negligible. The CSSM model includes pressure as a parameter in its expression for bulk modulus, while the present theory includes pressure, shear stress and specific volume. Shear stress enters its bulk, cross and shear moduli at non-isotropic states solely through the shear stress ratio, $q/p$. Throughout the elastic region this ratio lies below the value that mobilizes friction at critical state ($q/p' < M$). Further research to establish the intertheoretic relations for shear stress ratios below critical state value ($0 < q/p' < M$) is clearly needed.

The expressions for bulk modulus listed in Table 1 include a term that can be identified as the measure of a material's softness. The product of the contraction-swelling index and the uniform bulk index ($\kappa k$ or ${\kappa }^*k$) is a material constant that relates differential to uniform stiffness. A material with a vanishingly small value is hard. A material with a higher value than another material is softer than the other. That is, this product locates its material along a spectrum from hard to soft condensed matter.

\section{Concluding Remarks}

The multi-scaling theory of elasticity described here is a constraint theory that includes specific volume as an internal state variable and allows it to change between multi-scale equilibrium states at rates independent of the rate of applied stress. Its scope includes condensed matter in general and geomaterials in specific, ranging from porous solids to aggregates.

The theory supports the more commonly accepted empirical models for soils and conserves energy in closed loading cycles within the elastic region well away from failure. Its uniformity surface partitions the stress sub-space in the vicinity of the current state into contraction and swelling sub-domains and identifies the locus of states that are reachable without changes in packing. The major symmetry of the macroscopic elasticity and compliance tensors follows from equilibrium at macroscopic and mesoscopic scales. A contraction-swelling curve describes the locus of mesoscopic equilibrium packing pressures across the range of specific volumes.

The theory requires at least three coefficients to describe an isotropic material: its bulk, shear and contraction-swelling indices. It extrapolates the empirical expressions for bulk, cross and shear moduli established at isotropic loading states across the domain of state space. The theory includes the modern theory of elasticity as its regular limit and offers refinements to the Critical State Soil Mechanics model.

\acknowledgements

Professor Jitendrapal Sharma suggested the term contraction to describe compressive differential deformation of the interstitial phase. Dr. Alireza Najma checked the derivations and assisted with the data retrieval and presentation. Nurida Fatoullaeva assisted in the proof-reading of this paper. This research did not receive any specific grant from funding agencies in the public, commercial, or not-for-profit sectors.

\appendix
\section*{Packing Energy Examples}

A contraction-swelling relation in $\nu -ln\phi $ space takes the form:

\begin{tabular}{p{0.5in}p{3.3in}p{0.5in}}
 & $\delta \nu =\ -\ \xi \ \delta \phi /\phi $ & (A1) \\
\end{tabular}

\noindent
where $\xi $ denotes its tangential slope. The corresponding relation between packing pressure and specific volume is given by

\begin{tabular}{p{0.5in}p{3.3in}p{0.5in}}
 & $\beta =\ \phi -\ {\phi }_re^{\left\{\frac{{\nu }_r-\nu }{\xi }\right\}}=0$ & (A2) \\
\end{tabular}

\noindent
where ${\phi }_r$ denotes the reference packing pressure at reference specific volume (${\nu }_r$). The packing energy increment follows from Eqs. (7) and (A2)

\begin{tabular}{p{0.5in}p{3.3in}p{0.5in}}
 & $\delta P\left(\nu \right)=\ -\ {\phi }_re^{\left\{\frac{{\nu }_r-\nu }{\xi }\right\}}\delta \nu =\xi \ \delta \phi $ & (A3) \\
\end{tabular}

\noindent
Integrating Eq. (A3) as a function of specific volume yields

\begin{tabular}{p{0.5in}p{3.3in}p{0.5in}}
 & $P\left(\nu \right)=\xi (\phi -\ {\phi }_r)$ & (A4) \\
\end{tabular}

\noindent
Selecting a fully dispersed state as the reference state [62], yields

\begin{tabular}{p{0.5in}p{2.2in}p{1.6in}}
 & $P\left(\nu \right)=\xi \ \phi $ & $\ {\phi }_r\ \to 0,\ {\nu }_r\ \to \infty $ (A5) \\
\end{tabular}

\noindent
The contraction-swelling index in  $\nu -ln\phi $ space is related to the index in  $\nu -ln p_o/p_r $ space through Eqs. (100) and (132):

\begin{tabular}{p{0.5in}p{3.3in}p{0.5in}}
 & $\xi = \kappa \ /\ [2 - n - (a - 1)\kappa/\nu]$ & (A6) \\
\end{tabular}

\noindent
Note that for a semi-logarithmic constitutive relation, the index for one scale is a function of specific volume for the other scale.

The contraction-swelling relation in $ln\nu -ln\phi $ space takes the form:

\begin{tabular}{p{0.5in}p{3.3in}p{0.5in}}
 & ${\delta \nu }/{\nu }=\ -\ {\xi }^*{\delta \phi }/{\phi }$ & (A7) \\
\end{tabular}

\noindent
where ${\xi }^*$ denotes the tangential slope. The corresponding relation between packing pressure and specific volume is given by

\begin{tabular}{p{0.5in}p{3.3in}p{0.5in}}
 & $\beta =\ \phi -\ {\phi }_r{\left(\frac{{\nu }_r}{\nu }\right)}^{\left\{\frac{1}{\xi *}\right\}}=0$ & (A8) \\
\end{tabular}

\noindent
The specific packing energy increment follows from Eqs. (7) and (A8)

\begin{tabular}{p{0.5in}p{3.3in}p{0.5in}}
 & $\delta P\left(\nu \right)=\ -\ {\phi }_r{\left(\frac{{\nu }_r}{\nu }\right)}^{\left\{\frac{1}{\xi *}\right\}}\delta \nu =\nu \ {\xi }^{*\ }\delta \phi $ & (A9) \\
\end{tabular}

\noindent
Integrating Eq. (A9) as a function of specific volume yields

\begin{tabular}{p{0.5in}p{3.3in}p{0.5in}}
 & $P\left(\nu \right)=\ \left[\frac{{\xi }^*}{1-{\xi }^*}\right]{\phi }_r{\nu }_r\left[{\left(\frac{{\nu }_r}{\nu }\right)}^{\left\{\frac{{1-\xi }^*}{{\xi }^*}\right\}}-1\right]$ \\
 & $\ \ \ \ \ \ \ \ \ =\ \left[\frac{{\xi }^*}{1-{\xi }^*}\right]{\phi }_r{\nu }_r\left[{\left(\frac{\phi }{{\phi }_r}\right)}^{\left\{1-{\xi }^*\right\}}-1\right]$ & (A10) \\
\end{tabular}

\noindent
The contraction-swelling index in  $ln\nu -ln\phi $ space is related to the index in  $ln\nu -ln p_o/p_r $ space through Eqs. (102) and (132):

\begin{tabular}{p{0.5in}p{3.3in}p{0.5in}}
 & ${\xi }^* = {\kappa }^* \ /\ [2 - n - (a - 1){\kappa }^* ]$ & (A11) \\
\end{tabular}

\noindent
Note that for a logarithmic-logarithmic constitutive relation, the index for one scale is not a function of specific volume for the other scale.

\noindent

\noindent \textbf{\eject }

\noindent \textbf{Figure Captions}

\noindent 1 -- Centric Deformations

\noindent 2 -- Straining of Solid and Interstitial Phases

\noindent 3 -- Changes in Packing Pressure

\noindent 4 -- Contraction-Swelling Constitutive Relation

\noindent 5 -- Uniformity Surface through the Current State

\noindent 6 -- Internal Energy

\noindent 7 -- Complementary Internal Energy

\noindent 8 -- Uniformity Surface for an Isotropic Material

\noindent 9 -- Multi-Scale and Modern Theories of Elasticity

\noindent 10 -- Dense Fine Ottawa Sand under Isotropic Unloading ($D_r=75\%,\ k{\kappa }^*=0.64$) (after Dakoulas et al. 1992 -- reproduced with permission from ASCE)

\noindent 11 -- Loose Fine Ottawa Sand under Isotropic Unloading ($D_r=30\%,\ k{\kappa }^*=0.64$) (after Dakoulas et al. 1992 -- reproduced with permission from ASCE)

\noindent 12 -- Volume Changes in TDA (tire chips) under saturated, drained, isotropic compression (Wartman et al. 2007 -- reproduced with permission from ASCE)


\begin{thebibliography}{65}
\bibitem{Woodcock}
N. Woodcock
Geology and Environment In Britain and Ireland, 1 edition.
London: CRC Press, 1994.

\bibitem{Zytynski}
M. Zytynski, M. F. Randolph, R. Nova, and C. P. Wroth
``On modelling the unloading-reloading behaviour of soils,'' International Journal for Numerical and Analytical Methods in Geomechanics, vol. 2, no. 1, pp. 87--93, Jan. 1978.

\bibitem{Randolph}
M. F. Randolph, J. P. Carter, and C. P. Wroth,
``Driven piles in clay---the effects of installation and subsequent consolidation,''
G\'{e}otechnique, vol. 29, no. 4, pp. 361--393, Dec. 1979.

\bibitem{Houlsby}
G. T. Houlsby,
``The use of a variable shear modulus in elastic-plastic models for clays,''
Computers and Geotechnics, vol. 1, no. 1, pp. 3--13, Jan. 1985.

\bibitem{Wood}
D. M. Wood,
Soil Behaviour and Critical State Soil Mechanics, 1 edition.
Cambridge England; New York: Cambridge University Press, 1991.

\bibitem{Borja}
R. I. Borja, C.-H. Lin, and F. J. Mont\'{a}ns,
``Cam-Clay plasticity, Part IV: Implicit integration of anisotropic bounding surface model with nonlinear hyperelasticity and ellipsoidal loading function,''
Computer Methods in Applied Mechanics and Engineering, vol. 190, no. 26--27, pp. 3293--3323, 2001.

\bibitem{HoulsbyPuzrin}
G. T. Houlsby and A. M. Puzrin,
Principles of hyperplasticity: an approach to plasticity theory based on thermodynamic principles.
London: Springer, 2006.

\bibitem{Coombs}
W. M. Coombs and R. S. Crouch,
``Algorithmic issues for three-invariant hyperplastic Critical State models,''
Computer Methods in Applied Mechanics and Engineering, vol. 200, no. 25--28, pp. 2297--2318, Jun. 2011.

\bibitem{Krabbenhoft}
K. Krabbenhoft and A. V. Lyamin,
``Computational Cam clay plasticity using second-order cone programming,''
Computer Methods in Applied Mechanics and Engineering, vol. 209--212, pp. 239--249, 2012.

\bibitem{Stickle}
M. M. Stickle, P. De La Fuente, C. Oteo, M. Pastor, and P. Dutto,
``A modelling framework for marine structure foundations with example application to vertical breakwater seaward tilt mechanism under breaking wave loads,''
Ocean Engineering, vol. 74, pp. 155--167, 2013.

\bibitem{Golchin}
A. Golchin and A. Lashkari, ``A critical state sand model with elastic--plastic coupling,'' International Journal of Solids and Structures, vol. 51, no. 15--16, pp. 2807--2825, 2014.

\bibitem{Hong}
P. Y. Hong, J. M. Pereira, Y. J. Cui, A. M. Tang, F. Collin, and X. L. Li, ``An elastoplastic model with combined isotropic--kinematic hardening to predict the cyclic behavior of stiff clays,'' Computers and Geotechnics, vol. 62, pp. 193--202, 2014.

\bibitem{Pense}
S. Le Pense, ``Mean stress dependent nonlinear hyperelasticity coupled with damage stiffness degradation. A thermodynamical approach,'' Mechanics Research Communications, vol. 60, pp. 85--89, 2014.

\bibitem{Wong}
K. S. Wong and D. Ma\v{s}\'{i}n, ``Coupled hydro-mechanical model for partially saturated soils predicting small strain stiffness,'' Computers and Geotechnics, vol. 61, pp. 355--369, 2014.

\bibitem{Duriez}
J. Duriez and \'{E}. Vincens, ``Constitutive modelling of cohesionless soils and interfaces with various internal states: An elasto-plastic approach,'' Computers and Geotechnics, vol. 63, pp. 33--45, 2015.

\bibitem{Robin}
V. Robin, A. A. Javadi, O. Cuisinier, and F. Masrouri, ``An effective constitutive model for lime treated soils,'' Computers and Geotechnics, vol. 66, pp. 189--202, 2015.

\bibitem{Martinelli}
M. Martinelli, A. Burghignoli, and L. Callisto, ``Dynamic response of a pile embedded into a layered soil,'' Soil Dynamics and Earthquake Engineering, vol. 87, pp. 16--28, 2016.

\bibitem{HongPereira}
P. Y. Hong, J. M. Pereira, A. M. Tang, and Y. J. Cui, ``A two-surface plasticity model for stiff clay,'' Acta Geotech., vol. 11, no. 4, pp. 871--885, Aug. 2016.

\bibitem{Lloret}
M. Lloret-Cabot, S. W. Sloan, D. Sheng, and A. J. Abbo, ``Error behaviour in explicit integration algorithms with automatic substepping,'' International Journal for Numerical Methods in Engineering, vol. 108, no. 9, pp. 1030--1053, 2016.

\bibitem{Sternik}
K. Sternik, ``Elasto-plastic Constitutive Model for Overconsolidated Clays,'' Int J Civ Eng, vol. 15, no. 3, pp. 431--440, May 2017.

\bibitem{Prashant}
A. Prashant, D. Bhattacharya, and S. Gundlapalli, ``Stress-state dependency of small-strain shear modulus in silty sand and sandy silt of Ganga,'' G\'{e}otechnique, vol. 69, no. 1, pp. 42--56, Feb. 2018.

\bibitem{Shi}
Z. Shi, G. Buscarnera, and R. J. Finno, ``Simulation of cyclic strength degradation of natural clays via bounding surface model with hybrid flow rule,'' International Journal for Numerical and Analytical Methods in Geomechanics, vol. 42, no. 14, pp. 1719--1740, 2018.

\bibitem{Vrakas}
A. Vrakas, ``On the computational applicability of the modified Cam-clay model on the `dry' side,'' Computers and Geotechnics, vol. 94, pp. 214--230, Feb. 2018.

\bibitem{Yu}
H.-S. Yu, P.-Z. Zhuang, and P.-Q. Mo, ``A unified critical state model for geomaterials with an application to tunnelling,'' Journal of Rock Mechanics and Geotechnical Engineering, Dec. 2018.

\bibitem{Zhang}
Z. Zhang, Y. Chen, and Z. Huang, ``A novel constitutive model for geomaterials in hyperplasticity,'' Computers and Geotechnics, vol. 98, pp. 102--113, 2018.

\bibitem{Kang}
X. Kang and H. Liao, ``Bounding surface plasticity model for jointed soft rocks considering overconsolidation and structural decay,'' Computers and Geotechnics, vol. 108, pp. 295--307, Apr. 2019.

\bibitem{Nguyen}
H. H. Nguyen, H. Khabbaz, and B. Fatahi, ``A numerical comparison of installation sequences of plain concrete rigid inclusions,'' Computers and Geotechnics, vol. 105, pp. 1--26, Jan. 2019.

\bibitem{Silvestri}
V. Silvestri and C. Tabib, ``An Enhanced Solution for the Expansion of Cylindrical Cavities in Modified Cam Clay,'' in Advances in Numerical Methods in Geotechnical Engineering, 2019, pp. 101--117.

\bibitem{Oden}
J. T. Oden, ``Simulation-Based Engineering Science: A National Science Foundation Blue Ribbon Report,'' 03-May-2006. [Online]. Available: https://www.nsf.gov/pubs/reports/sbes\_final\_report.pdf.

\bibitem{Batterman}
R. Batterman, ``The Tyranny of Scales,'' The Oxford Handbook of Philosophy of Physics, Feb. 2013.

\bibitem{Love}
A. E. H. Love, A Treatise on the Mathematical Theory of Elasticity, 4th Revised ed. edition. New York: Dover Publications, 2011.

\bibitem{Todhunter}
I. Todhunter and K. Pearson, A history of the theory of elasticity and of the strength of materials: from Galilei to Lord Kelvin, 2 vols. New York: Dover Publications, 1960.

\bibitem{Schofield}
A. N. Schofield and C. P. Wroth, Critical State Soil Mechanics. Maidenhead, UK: McGraw-Hill, 1968.

\bibitem{Bolton}
M. Bolton, Guide to Soil Mechanics. London: Macmillan, 1979.

\bibitem{Atkinson}
J. H. Atkinson and P. L. Bransby, Mechanics of Soils: An Introduction to Critical State Soil Mechanics. London?; New York: McGraw-Hill Inc.,US, 1978.

\bibitem{Butterfield}
R. Butterfield, ``A natural compression law for soils (an advance on e--log p','' G\'{e}otechnique, vol. 29, no. 4, pp. 469--480, Dec. 1979.

\bibitem{Hashiguchi}
K. Hashiguchi, ``On the linear relations of V--ln p and ln v--ln p for isotropic consolidation of soils,'' International Journal for Numerical and Analytical Methods in Geomechanics, vol. 19, no. 5, pp. 367--376, 1995.

\bibitem{Hardin}
B. O. Hardin and W. L. Black, ``Vibration Modulus of Normally Consolidated Clay,'' Journal of the Soil Mechanics and Foundations Division, vol. 94, no. 2, pp. 353--370, 1968.

\bibitem{Shibuya}
S. Shibuya, S. C. Hwang, and T. Mitachi, ``Elastic shear modulus of soft clays from shear wave velocity measurement,'' G\'{e}otechnique, vol. 47, no. 3, pp. 593--601, Jun. 1997.

\bibitem{Vardanega}
Vardanega P. J. and Bolton M. D., ``Stiffness of Clays and Silts: Normalizing Shear Modulus and Shear Strain,'' Journal of Geotechnical and Geoenvironmental Engineering, vol. 139, no. 9, pp. 1575--1589, Sep. 2013.

\bibitem{Collins}
Collins I. F. and Houlsby G. T., ``Application of thermomechanical principles to the modelling of geotechnical materials,'' Proceedings of the Royal Society of London. Series A: Mathematical, Physical and Engineering Sciences, vol. 453, no. 1964, pp. 1975--2001, Sep. 1997.

\bibitem{HoulsbyAmorosi}
G. T. Houlsby, A. Amorosi, and E. Rojas, ``Elastic moduli of soils dependent on pressure: a hyperelastic formulation,'' G\'{e}otechnique, vol. 55, no. 5, pp. 383--392, 2005.

\bibitem{Maugin}
G. A. Maugin, The Thermomechanics of Plasticity and Fracture, 1st edition. Cambridge England; New York: Cambridge University Press, 1992.

\bibitem{KestinRice}
J. Kestin and J. R. Rice, ``Paradoxes in the Application of Thermodynamics to Strained Solids,'' in A Critical Review of Thermodynamics, E. B. Stuart, A. J. Brainard, and B. Gal-Or, Eds. Baltimore: Mono Book Company, 1970, pp. 275--298.

\bibitem{Hueckel}
T. Hueckel and G. Maier, ``Incremental boundary value problems in the presence of coupling of elastic and plastic deformations: A rock mechanics oriented theory,'' International Journal of Solids and Structures, vol. 13, no. 1, pp. 1--15, Jan. 1977.

\bibitem{Collins2002}
I. F. Collins, ``Associated and Non-Associated Aspects of the Constitutive Laws for Coupled Elastic/Plastic Materials,'' International Journal of Geomechanics, vol. 2, no. 2, pp. 259--267, 2002.

\bibitem{Kestin}
J. Kestin, ``Local-equilibrium formalism applied to mechanics of solids,'' International Journal of Solids and Structures, vol. 29, no. 14, pp. 1827--1836, Jan. 1992.

\bibitem{Biot}
M. A. Biot, ``Theory of Stress-Strain Relations in Anisotropic Viscoelasticity and Relaxation Phenomena,'' Journal of Applied Physics, vol. 25, no. 11, pp. 1385--1391, Nov. 1954.

\bibitem{Rice}
J. R. Rice, ``Inelastic constitutive relations for solids: An internal-variable theory and its application to metal plasticity,'' Journal of the Mechanics and Physics of Solids, vol. 19, no. 6, pp. 433--455, Nov. 1971.

\bibitem{Zia}
R. K. P. Zia, E. F. Redish, and S. R. McKay, ``Making sense of the Legendre transform,'' American Journal of Physics, vol. 77, no. 7, pp. 614--622, Jun. 2009.

\bibitem{Szalwinski}
C. M. Szalwinski, ``The particle stress tensor,'' G\'{e}otechnique, vol. 33, no. 2, pp. 181--182, Jun. 1983.

\bibitem{Goddard}
Goddard J. D. and Enderby John Edwin, ``Nonlinear elasticity and pressure-dependent wave speeds in granular media,'' Proceedings of the Royal Society of London. Series A: Mathematical and Physical Sciences, vol. 430, no. 1878, pp. 105--131, Jul. 1990.

\bibitem{Richart}
J. F. E. Richart, J. J. R. Hall, and R. D. Woods, Vibrations of Soils and Foundations. Englewood Cliffs, N.J: Prentice Hall, 1970.

\bibitem{Dakoulas}
Dakoulas Panos and Sun Yuanhui, ``Fine Ottawa Sand: Experimental Behavior and Theoretical Predictions,'' Journal of Geotechnical Engineering, vol. 118, no. 12, pp. 1906--1923, Dec. 1992.

\bibitem{Yang}
L. Yang and L. Salvati, ``Small Strain Properties of Sands with Different Cement Types,'' International Conferences on Recent Advances in Geotechnical Earthquake Engineering and Soil Dynamics, May 2010.

\bibitem{Wartman}
Wartman Joseph, Natale Mark F., and Strenk Patrick M., ``Immediate and Time-Dependent Compression of Tire Derived Aggregate,'' Journal of Geotechnical and Geoenvironmental Engineering, vol. 133, no. 3, pp. 245--256, Mar. 2007.

\bibitem{Anderson}
O. L. Anderson, ``2 - Determination and Some Uses of Isotropic Elastic Constants of Polycrystalline Aggregates Using Single-Crystal Data,'' in Physical Acoustics, vol. 3, W. P. Mason, Ed. Academic Press, 1965, pp. 43--95.

\bibitem{Munro}
R. G. Munro, ``Effective Medium Theory of the Porosity Dependence of Bulk Moduli,'' Journal of the American Ceramic Society, vol. 84, no. 5, pp. 1190--1192, May 2001.

\bibitem{Krief}
M. Krief, J. Garat, J. Stellingwerff, and J. Ventre, ``A Petrophysical Interpretation Using the Velocities of P and S Waves (Full-Waveform Sonic),'' The Log Analyst, vol. 31, no. 06, Nov. 1990.

\bibitem{Pickett}
G. R. Pickett, ``Acoustic Character Logs and Their Applications in Formation Evaluation,'' Journal of Petroleum Technology, vol. 15, no. 06, pp. 659--667, Jun. 1963.

\bibitem{Knackstedt}
M. A. Knackstedt, C. H. Arns, and W. Val Pinczewski, ``Velocity--porosity relationships: Predictive velocity model for cemented sands composed of multiple mineral phases,'' Geophysical Prospecting, vol. 53, no. 3, pp. 349--372, 2005.

\bibitem{Roscoe}
K. H. Roscoe, A. N. Schofield, and A. Thurairajah, ``Yielding of Clays in States Wetter than Critical,'' G\'{e}otechnique, vol. 13, no. 3, pp. 211--240, Sep. 1963.

\bibitem{RoscoeSchofield}
K. H. Roscoe, A. N. Schofield, and C. P. Wroth, ``On the Yielding of Soils,'' G\'{e}otechnique, vol. 8, no. 1, pp. 22--53, Mar. 1958.

\bibitem{Batterman2016}
R. Batterman, ``Intertheory Relations in Physics,'' in The Stanford Encyclopedia of Philosophy, Fall 2016., E. N. Zalta, Ed. Metaphysics Research Lab, Stanford University, 2016.

\bibitem{Jenike}
A. W. Jenike and R. T. Shield, ``On the Plastic Flow of Coulomb Solids Beyond Original Failure,'' Journal of Applied Mechanics, vol. 26, pp. 599--602, 1959.

\end{thebibliography}
\end{document}